\documentclass[preprint2]{aastex}
\newcommand{\tnin}{\mathrm{T}_{90}}
\shorttitle{Short Title Goes Here}
\shortauthors{Dib et al.}
\begin{document}
%% --------------------------------------------------------
\title{{\emph{RXTE}} Monitoring of the Anomalous X-ray Pulsar
1E~1048.1$-$5937: Long-Term Variability and the 2007 March Event}
%% --------------------------------------------------------
\author{Rim~Dib\altaffilmark{1},
        Victoria~M.~Kaspi\altaffilmark{1}, and
        Fotis~P.~Gavriil\altaffilmark{2,}\altaffilmark{3}}
%% --------------------------------------------------------
%% \email{}
%% --------------------------------------------------------
\altaffiltext{1}{Department of Physics, McGill University,
                 Montreal, QC H3A~2T8}
\altaffiltext{2}{Center for Research and Exploration in
Space Science and Technology, NASA Goddard Space Flight Center, Code 662, 
Greenbelt, MD 20771, USA.}
\altaffiltext{3}{Department of Physics, University of
Maryland Baltimore County, 1000 Hilltop Circle, Baltimore, MD 21250, USA.}
%% --------------------------------------------------------
\begin{abstract}
%% --------------------------------------------------------
After three years of no unusual activity, Anomalous X-ray Pulsar
1E~1048.1$-$5937 reactivated in 2007 March. We report on the detection of a
large glitch ($\Delta\nu/\nu$~=~1.63(2)~$\times$~10$^{-5}$) on 2007 March 26
(MJD 54185.9), contemporaneous with the onset of a pulsed-flux flare, the
third flare observed from this source in 10 years of monitoring with the
{\emph{Rossi X-ray Timing Explorer}}. Additionally, we report on a detailed
study of the evolution of the timing properties, the pulsed flux, and the
pulse profile of this source as measured by {\emph{RXTE}} from 1996 July to
2008 January. In our timing study, we attempted phase coherent timing of all
available observations. We show that in 2001, a timing anomaly of uncertain
nature occurred near the rise of the first pulsed flux flare; we show that a
likely glitch ($\Delta\nu/\nu$~=~2.91(9)~$\times$~10$^{-6}$) occurred in
2002, near the rise of the second flare, and we present a detailed
description of the variations in the spin-down. In our pulsed flux study, we
compare the decays of the three flares and discuss changes in the hardness
ratio. In our pulse profile study, we show that the profile exhibited large
variations near the peak of the first two flares, and several small
short-term profile variations during the most recent flare. Finally, we
report on the discovery of a small burst 27 days after the peak of the last
flare, the fourth burst discovered from this source. We discuss the
relationships between the observed properties in the framework of the
magnetar model.
%% --------------------------------------------------------
\end{abstract}
%% --------------------------------------------------------
%% --------------------------------------------------------
%% A small citation guide:
%% \citep{jon90} (Jones et al. 1990) [no comma] 
%% \cite(t){jon90} Jones et al. (1990) [no comma]
%%
%% \citep*{jon90} (Jones, Baker, and Williams 1990) [one arg, no comma]
%% \citet*{jon90} Jones, Baker, and Williams (1990) [one arg]
%%
%% \citealp(jon90) Jones et al. 1990  [no comma]
%% \citealt{jon90} Jones et al. 1990  [no comma, been using this]
%%
%% \citealp*{jon90} Jones, Baker, and Williams 1990 [all and no comma,this]
%% \citealt*{jon90} Jones, Baker, and Williams 1990 [all and no comma]
%% --------------------------------------------------------
%% --------------------------------------------------------
%% ---------- P A P E R -- S T A R T S -- H E R E ---------
%% --------------------------------------------------------
%% --------------------------------------------------------

\section{Introduction}

%% 1048 is an axp %%
The source 1E~1048.1$-$5937 is part of the class of sources known as
Anomalous X-ray Pulsars (AXPs). They have generally been characterized by a
persistent X-ray luminosity in excess of available spin-down power, although
there are exceptions (e.g. AXP 1E~1547.0$-$5408 in 2006 \citep{csh+07,gg07}).
AXPs are young, isolated pulsars with a large inferred magnetic field
($>$~10$^{14}$~G). They are detected across the electromagnetic spectrum
from the radio (in 2 cases) to the hard X-ray regime. Just like a closely
related class of pulsars, the Soft Gamma Repeaters (SGRs), AXPs exhibit a
wide range of variability, including but not limited to spectral
variability, timing glitches, X-ray bursts, X-ray pulsed and persistent flux
``flares'', and pulse profile changes. For recent reviews, see \cite{k07} and
\cite{m08}.

%% magnetar model %%
The magnetar model \citep{td95,td96a,tlk02} recognizes the power source of
these objects to be the decay of their strong magnetic fields. In this
model, the bursts of high-energy emission are thought to occur when the
crust succumbs to the internal magnetic stresses and deforms. The
deformation twists the footpoints of the external magnetic field, driving
currents into the magnetosphere and twisting it relative to the standard
dipolar geometry. These magnetospheric currents resonantly cyclotron-scatter
seed surface thermal photons, giving rise to the non-thermal component of the
spectrum, usually fitted to a power-law model below 10~keV. Additionally,
the high energy X-ray spectrum of magnetars may be explained by the
existence of a plasma corona contained within the closed magnetosphere
\citep{BT07}.

%% 1048 previous work %%
We have been monitoring 1E~1048.1$-$5937 with the {\emph{Rossi X-ray Timing
Explorer}} ({\em{RXTE}}) since 1997. During that time, the AXP has exhibited
significant timing and pulsed flux variability. Early regular monitoring
showed that the spin-down of 1E~1048.1$-$5937 was so unstable that phase
coherence could be maintained for periods of only a few months at a time
\citep{kgc+01}. In late 2001, two small bursts were detected from this AXP
\citep{gkw02}. The first of the two bursts coincided with the rise of the
first of two consecutive slow pulsed flux flares \citep{gk04}. The second
flare, the longer-lasting of the two, decayed during the second half of
2002, and throughout 2003 and 2004. A third burst was observed from the
source during this decay \citep{gkw06}. While the second pulsed flux flare
was ongoing, \cite{mts+04} and \cite{tmt+05} reported an enhancement in the
total flux of the source followed by a decay based on data from X-ray
imaging observations. The source was also seen to brighten in the IR at the
onset of the second flare \citep{wc02,ics+02}.

%%% 1048 after the flares %%
%%% The spin-down IN PREVIOUS PAPER was determined in short intervals 
%%% by calculating the slope of
%%% three consecutive frequency measurements. Each frequency measurement was
%%% obtained by phase-connecting a group of three closely spaced observations
%%% (see Section~\ref{sec:observations}. 
%%% 1048 after the flares %%
In 2003, during the decay of the second flare, \cite{gk04} reported
order-of-magnitude variations in the spin-down of the pulsar on timescales
of weeks to months.  In 2004, near the end of the decay of the second flare,
the source entered a quiescent period in which the pulsed flux slowly
decreased, with much smaller and more monotonic variations in the spin-down.
Then, in 2007 March, the source entered a new active phase.
\cite{dkgw07UPDATED} reported the detection of a sudden spin-up accompanied
by pulsed flux increase (hereafter referred to as the third flare) in
regular {\em{RXTE}} monitoring data. The enhancement in the phase-averaged X-ray
and infrared fluxes that accompanied this new flare are discussed
in detail in \cite{tgd+08}. \cite{wbk+08} reported on an optical
enhancement, and very recently \cite{dml+09} have reported
contemporaneous optical pulsations.

%% reporting %%
Here we present a detailed analysis of all {\em{RXTE}} observations of
1E~1048.1$-$5937 that were taken between 1996 July~03 and 2008 January~09.
We report the results of an in-depth analysis of the timing behavior, pulsed
flux changes, and pulse profile variations. These results include but are
not limited to those obtained from the analysis of the 2007 March events. We
also report on the detection of a fourth small burst on 2007 April 28.  Our
observations are described in Section~\ref{sec:observations}. Our timing,
pulsed morphology, and pulsed flux analyses are presented, respectively, in
Sections~\ref{sec:timing}, \ref{sec:profiles}, and~\ref{sec:flux}. In
Section~\ref{sec:bursts}, we discuss the most recent burst. Finally, in
Section~\ref{sec:discussion}, we compare the observed properties of
1E~1048.1$-$5937 to those of the other AXPs, and we discuss the implications
of our findings in the framework of the magnetar model.

\section{Observations}
\label{sec:observations}

The results presented here were obtained using the proportional counter
array (PCA) on board {\em{RXTE}}. The PCA consists of an array of five
collimated xenon/methane multi-anode proportional counter units (PCUs)
operating in the 2$-$60 keV range, with a total effective area of
approximately 6500~cm$^2$ and a field of view of $\sim$1$^{\circ}$ FWHM
\citep{jsg+96}.

%%% [[LINEBREAK ASTROPH]].
There are 841 {\em{RXTE}} observations of \linebreak
1E~1048.1$-$5937 taken between
MJD~50294.3 (1996 July~03) and MJD~54474.7 (2008 January~09). We used 821
of them for the analysis presented in this paper. The remaining observations
were excluded for various reasons (unusually short observations, pointing
errors, or missing files).

\begin{figure*}
\begin{center}
\includegraphics[scale=0.95]{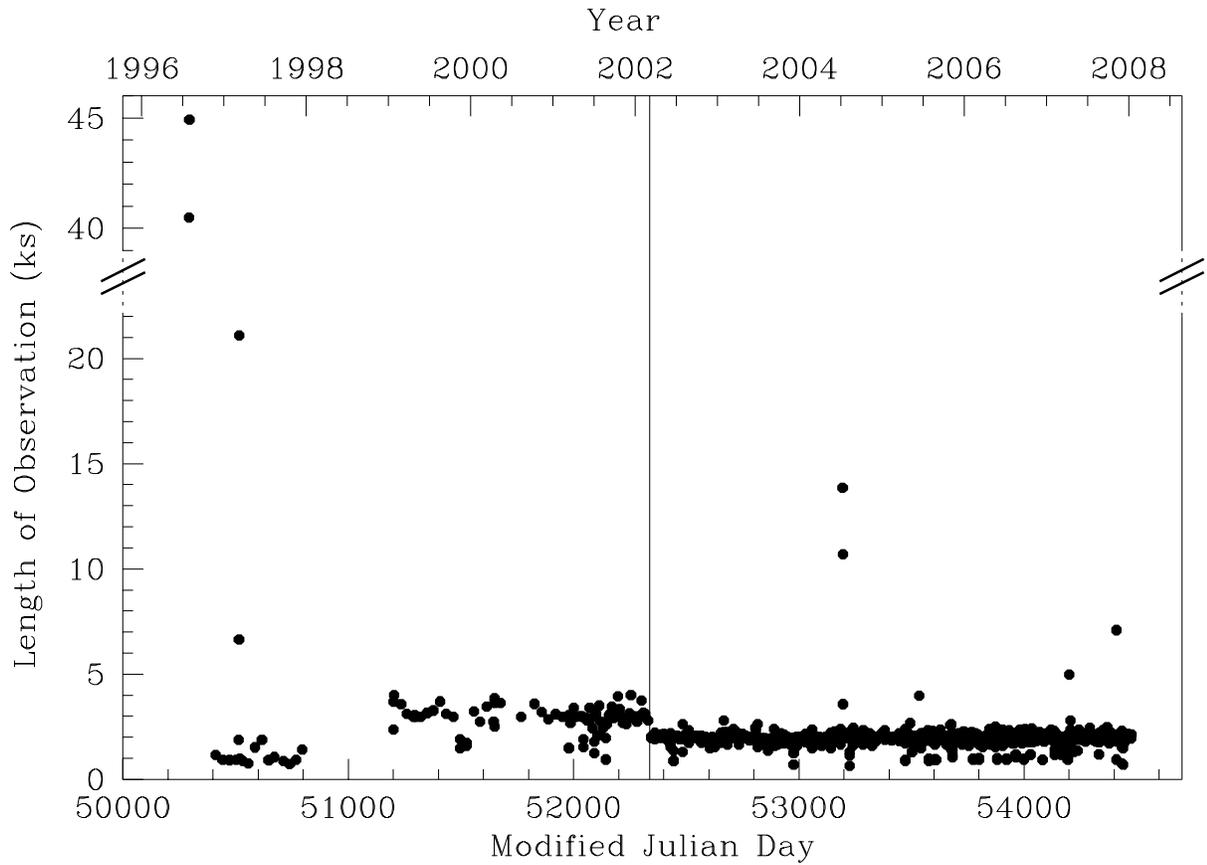}
\caption
{Length (on-source integration time) of the {\emph{RXTE}} observations of
1E~1048.1$-$5937 used in this paper versus epoch. The solid line indicates
when we adopted the strategy of observing the source with sets of three
closely spaced observations.
\label{plot-obslen}}
\end{center}
\end{figure*}

\begin{figure*}
\begin{center}
\includegraphics[scale=0.95]{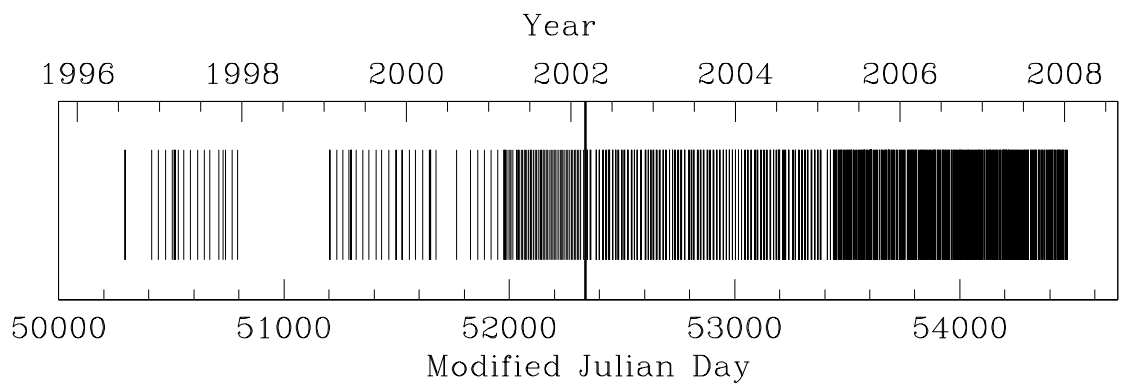}
\caption
{Epochs of the observations of 1E~1048.1$-$5937 used in this paper. The
bold line indicates  when we
adopted the strategy of observing the source with sets of three
closely spaced observations.
\label{plot-obstime}}
\end{center}
\end{figure*}

The length of the observations varied between 0.75~ks and 45~ks, but most of
them were 2~ks long (see Figure~\ref{plot-obslen}). The time intervals
between the observations are shown in Figure~\ref{plot-obstime}. The
observation frequency varied over the years from once per month to several
times per month. Because it was difficult to achieve long-term
phase-coherent timing for this source \citep{gk04}, in 2002 March, we
adopted the strategy of observing it three times every two weeks with three
closely spaced observations. The bold vertical line in
Figures~\ref{plot-obslen} and~\ref{plot-obstime} mark when this strategy was
implemented. The observing frequency increased to three times per week in
2005 March.

Throughout the monitoring, we used the {\tt GoodXenonwithPropane} data mode
to observe this source, except during {\em{RXTE}} Cycles~10 and~11 when we
used the {\tt GoodXenon} mode.  Both data modes record photon arrival times
with 1-$\mu$s resolution and bin photon energies into one of 256 channels.
To maximize the signal-to-noise ratio, we analysed only those events from
the top Xenon layer of each PCU.

\section{Phase-Coherent Timing Study: Analysis and Results}
\label{sec:timing}

\subsection{Long-Term Timing}
\label{sec:timf0}

To do the timing analysis, photon arrival times at each epoch were adjusted
to the solar system barycenter.  Resulting arrival times were binned with
31.25-ms time resolution. In the timing timing analysis, we included only
events in the energy range 2$-$5.5~keV, to maximize the signal-to-noise
ratio of the pulse. Each barycentric binned time series was epoch-folded
using an ephemeris determined iteratively by maintaining phase coherence as
we describe below. When an ephemeris was not available, we folded the time
series using a frequency obtained from a periodogram.  Resulting pulse
profiles, with 64 phase bins, were cross-correlated in the Fourier domain
with a high signal-to-noise template created by adding phase-aligned
profiles. The cross-correlation returned an average pulse time of arrival
(TOA) for each observation corresponding to a fixed pulse phase. The pulse
phase $\phi$ at any time $t$ can usually be expressed as a Taylor expansion,

\begin{equation}
\label{eq:polynomials}
\phi(t) = \phi_{0}(t_{0})+\nu_{0}(t-t_{0})+
\frac{1}{2}\dot{\nu_{0}}(t-t_{0})^{2}
+\frac{1}{6}\ddot{\nu_{0}}(t-t_{0})^{3}+{\ldots},
\end{equation}

\noindent where $\nu$~$\equiv$~1/$P$ is the pulse frequency,
$\dot{\nu}$~$\equiv$~$d\nu$/$dt$, etc$.$, and subscript ``0'' denotes a
parameter evaluated at the reference epoch $t=t_0$.

To obtain ephemerides for data prior to 2001, we fitted the TOAs to the
above polynomial using the pulsar timing software package \texttt{TEMPO}\footnote{See
http://www.atnf.csiro.au/research/pulsar/tempo.}. \texttt{TEMPO} also returned an
absolute pulse number associated with each TOA, corresponding to the number
of times that the pulsar rotated since the first TOA. Since the spin-down of
this source was unstable, phase coherence could only be maintained for
periods of several months at a time \citep{kgc+01}.

After 2002 March~02, we started observing 1E~1048.1$-$5937 using sets of
three closely spaced observations. For data after this date, we adopted a
new timing strategy. We broke the list of TOAs into several segments. Each
segment lasted between 8 and 16 weeks (4 to 8 weeks after 2005 March), with
an overlap between one segment and the next of at least four weeks (2 weeks
after 2005 March), except for the week of 2003 April 13 (MJD 52742) where
the overlap was of two weeks only, and except at the onset of the flares
where there was no overlap. For each two overlapping segments, we used \texttt{TEMPO}
to fit the TOAs with Equation~\ref{eq:polynomials} and extract pulse
numbers. We then checked that the pulse numbers of the observations present
in both segments were the same. This gave us confidence that the two
overlapping ephemerides were consistent with each other and that phase
coherence was not lost. Combining all overlapping segments between two given
dates yielded a time series of absolute pulse number versus TOA. The errors
on the TOAs were converted into fractional errors on the pulse numbers. We
also used \texttt{TEMPO} to fit the TOAs obtained between 2001 and 2002 March 02 with
two non-overlapping ephemerides and extracted pulse numbers.

All the pulse numbers obtained using the procedure above were then organized
into four different pulse number versus TOA time series: a time series
covering the time interval between 2001 and the onset of the first flare
(2001 March to 2001 October), a time series covering the interval between
the onset of the first flare and that of the second (2001 November to 2002
April), a time series covering the interval between the onset of the second
flare and that of the third (2002 May to 2007 March), and a time series
covering the interval between the onset of the third flare and the date of
the last observation included in this paper (2007 March to 2008 January).

Because of the instability of the spin-down of AXP 1E~1048.1$-$5937, timing
solutions spanning long periods of time required the use of very high-order
polynomials which tended to oscillate at the end points of fitted intervals.
Instead of using these polynomials, we used splines. A spline is a
piecewise polynomial function. It consists of polynomial pieces of degree
$n$ (here $n$~=~5) defined between points called knots. The two polynomial
pieces adjacent to any knot share a common value and common derivative
values at the knot, through the derivative of order $n$$-$2 (see
\citeauthor{dierckx}, 1975 for more details about splines). We fit a spline
function through each of the above time series, weighted by the inverse of
the square of the fractional errors on the pulse numbers.  To minimize
oscillations in the spline due to noise, we set the spline smoothing
parameter to allow the RMS phase residual obtained after subtracting the
spline from the data points to be twice the average 1$\sigma$ uncertainty in
the pulse phase. The smoothing parameter controls the tradeoff between
closeness and smoothness of fit by varying the polynomial coefficients and
the spacing between the knots. We found the uncertainties on the spline by
adding Gaussian noise to our data points 500 times, with mean equal to the
1$\sigma$ uncertainty on each data point, fitting each time with a spline,
averaging all the splines, and finding the standard deviation at each point.

The derivative of the spline function is the frequency of the pulsar, and
the second derivative of the spline function is the frequency derivative of
the pulsar.

\begin{figure*}
\begin{center}
\includegraphics[scale=0.6]{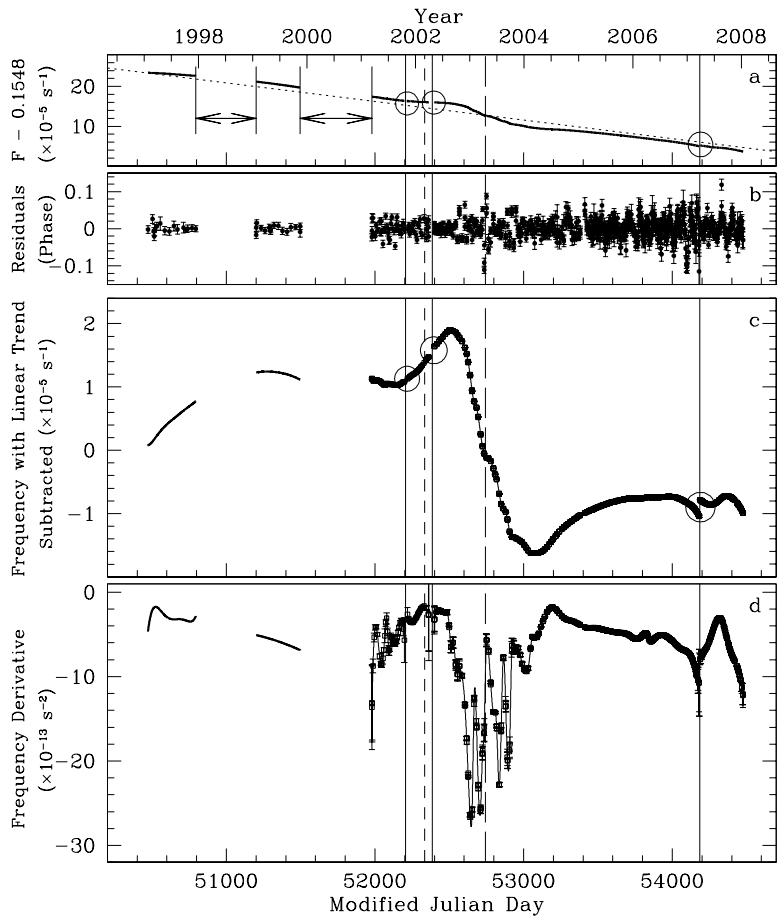}
%%% \caption {Timing properties of AXP 1E~1048.1$-$5937. {\bf{Panel~a:}}
%%% Long-term evolution of the frequency of 1E~1048.1$-$5937.
%%% The slope of the diagonal dotted line is the average spin-down of the pulsar
%%% ($\sim$$-$5.4~$\times$~10$^{-13}$~s$^{-2}$). The deviations from the average
%%% spin-down are clear to the eye. The three circles are centered at the start
%%% of the three pulsed flux flares, and mark the location of a timing anomaly,
%%% a likely glitch, and a glitch. {\bf{Panel~b:}} Timing residuals obtained
%%% after subtracting the TOAs from the ephemerides plotted in panel~a.
%%% {\bf{Panel~c:}} Long-term frequency evolution with the long-term average
%%% spin-down subtracted. For the first two plotted curves, the error bars are
%%% smaller than the width of the lines. The three circles are centered at the
%%% start of the three pulsed flux flares (see Figures~\ref{plot-flux}
%%% and~\ref{plot-compare}), and mark the location of a timing anomaly and two
%%% glitches (see Sections~\ref{sec:timf1}$-$\ref{sec:timf3} in the text).
%%% {\bf{Panel~d:}} Long-term evolution of the frequency derivative of the
%%% pulsar. {\bf{All panels:}} The three solid lines indicate the onset of the
%%% three pulsed flux flares. The short-dashed line marks the epoch when we
%%% started observing the source in sets of three closely spaced observations.
%%% The long-dashed line indicates the epoch when the overlap in the partial
%%% ephemerides was for a single set of three observations (see
%%% Section~\ref{sec:timf0} for details). 
\caption {Timing properties of AXP 1E~1048.1$-$5937. ({\emph{a}})
Long-term evolution of the frequency of 1E~1048.1$-$5937.
The slope of the diagonal dotted line is the average spin-down of the pulsar
($\sim$$-$5.4~$\times$~10$^{-13}$~s$^{-2}$). The deviations from the average
spin-down are clear to the eye. The three circles are centered at the start
of the three pulsed flux flares (see Figures~\ref{plot-flux}
and~\ref{plot-compare}), and mark the location of a timing anomaly,
a likely glitch, and a glitch. ({\emph{b}}) Timing residuals obtained
after subtracting the TOAs from the ephemerides plotted in panel~a.
({\emph{c}}) Long-term frequency evolution with the long-term average
spin-down subtracted. For the first two plotted curves, the error bars are
smaller than the width of the lines. The three circles are centered at the
start of the three pulsed flux flares, and mark the location of a timing anomaly and two
glitches (see Sections~\ref{sec:timf1}$-$\ref{sec:timf3} in the text).
({\emph{d}}) Long-term evolution of the frequency derivative of the
pulsar. {\emph{All panels:}} The three solid lines indicate the onset of the
three pulsed flux flares. The short-dashed line marks the epoch when we
started observing the source in sets of three closely spaced observations.
The long-dashed line indicates the epoch when the overlap in the partial
ephemerides was for a single set of three observations (see
Section~\ref{sec:timf0} for details).
\label{plot-timing}}
\end{center}
\end{figure*}

The results of this timing analysis are presented in
Figure~\ref{plot-timing}.  
%%% In all four panels of the Figure, the three solid
%%% vertical lines mark the onset of the three flares. The short-dashed line
%%% marks the epoch when we started observing the source in sets of three
%%% closely spaced observations. The long-dashed line marks the location where
%%% the pulsar timing behavior was so noisy that phase-connected segments could
%%% only overlap for a single set of three observations.
%% top panel %%
The top panel of Figure~\ref{plot-timing} shows frequency versus
time. The first horizontal double arrow indicates a time interval in which
1E~1048.1$-$5937 was not observed with {\em{RXTE}}. The second horizontal
arrow indicates a time interval in which data were so sparse that multiple
phase-coherent timing solutions could be found. The first two plotted curves
are ephemerides obtained using \texttt{TEMPO} only. They are consistent with the
first three ephemerides reported in \cite{kgc+01}. 
%%% The error bars are smaller than the width of the lines. 
The remaining plotted curves are ephemerides obtained from taking the
derivatives of spline functions. The slope of the diagonal dotted line is
the average spin-down of the pulsar. The deviations from the average
spin-down are clear to the eye.  
%%% The three circles are centered at the start of the three flares, and they
%%% mark the location of a timing anomaly and two glitches (see
%%% Sections~\ref{sec:timf1}, \ref{sec:timf2}, and~\ref{sec:timf3}).  
Note that since the onsets of the first two flares were accompanied by
significant pulse profile changes, we did not include the data from the two
weeks surrounding each in this Figure (see Sections~\ref{sec:timf1},
and~\ref{sec:timf2}).
%%% Although precision timing relies on profile stability, given the relatively
%%% subtle nature of the changes, they did not impact our timing results
%%% significantly most of the time. When this was not the case, we generated
%%% three sets of toas as explained above. This was done for the second flare as
%%% well.

%% second panel %%
The second panel of Figure~\ref{plot-timing} shows the timing residuals for
all phase-connected intervals. The RMS residuals for the two intervals fitted
with \texttt{TEMPO} are 2.4\% and 1.9\% of a pulse cycle. The RMS residuals for the
four intervals fitted with splines are 3.2\%, 3.5\%, 1.2\%, and 2.1\% of
the pulse cycle. The slow increase in the values of the uncertainties
between 2004 and the onset of the third flare reflects a decrease in 
signal-to-noise that is due both to a decrease in the pulsed flux of the source and
to a decrease in the effective number of operational PCUs onboard
{\emph{RXTE}}. The uncertainties are smaller after the flare due to the rise
in the pulsed flux.

%% third panel %%
The third panel of Figure~\ref{plot-timing} shows the frequency versus time
after having subtracted the long-term average linear trend shown in the top
panel. The two early ephemerides were obtained with \texttt{TEMPO}. The curves,
representing the remaining ephemerides, were obtained from derivatives of
splines. The points marked as squares are the values of the derivative
evaluated at the epochs at which observations were taken. A detailed timing
analysis of data inside the circles is done in Sections~\ref{sec:timf1},
\ref{sec:timf2}, and \ref{sec:timf3}.

%% third panel para 2 %% we say months.%%
We note in this panel that the spin-down of the pulsar was significantly
enhanced starting a few months after the onset of the second flare, a
phenomenon which lasted until 2004. In this period of time the pulsar's
rotational evolution became much noisier and phase-coherent timing would not
have been possible without the availability of sets of three closely spaced
observations. Note that \cite{gk04} previously reported this phenomenon
without using long-term phase-coherent timing: they obtained individual
frequency measurements by finding the frequency that best fit each three
observations.

%% third panel para 3 %%
In the same panel, we can also see that the pulsar then entered a quiescent
period in mid-2004 during which the frequency evolution was closer to the
long-term average.  Note that even though the frequency evolution in the
pre-2000 and post-2004 years looks stable, an analysis performed by
\cite{adl+08} of 1E~1048.1$-$5937 data between 2004 and 2007 reveals that
the amplitude of the timing noise (deviations from a simple spin-down) of
this AXP is significantly larger than that seen thus far for any other AXP.

%% third panel para 4 %%
Finally, in 2007 March, the pulsar underwent one of the largest glitches yet
observed in an AXP. The frequency jump inside the third circle is clear to
the eye. This is discussed in detail in Section~\ref{sec:timf3}.

%% 4th panel %%
The fourth panel of Figure~\ref{plot-timing} shows the frequency derivative
versus time. The first two plotted curves are obtained with \texttt{TEMPO}. The
remaining four curves, spanning data from 2001 to 2008, and separated by the
solid lines that mark the onset of the three flares, are obtained by taking
the second derivative of spline functions. The points marked as squares are
the values of that derivative evaluated at the epochs where observations
were taken. Note that the large error bars at the beginning and end of each
curve reflect the fact that the extremes of the curves are not well
constrained by the data.

%% 4th panel para 2 %% we say months %%
We note in this panel that, starting a few months after the onset of the
second flare, the pulsar underwent frequent and significant variations in
its spin-down on timescales of weeks to months. The variations noted in this
plot are consistent with those reported by \cite{gk04} although our analysis
here has higher time resolution. This is because in that study, the
spin-down was determined in short intervals by calculating the slope of
three consecutive frequency measurements, and each frequency measurement was
obtained by phase-connecting a group of three closely spaced observations.

%%% More specifically, we note that $\dot{\nu}$ was stable from 29 days
%%% preceding the peak of the second flare, until 41 days following the flare,
%%% fluctuating around the value $\sim$$-$2.3$\times10^{-13}$~s$^{-2}$ with
%%% variations on the order of 0.08$\times10^{-13}$~s$^{-2}$ every two weeks. Then,
%%% from the 41 days to 141 days after the flare, $\dot{\nu}$ dropped to
%%% $-$9.94$\times10^{-13}$~s$-2$ with an average drop of 2$\times10^{-13}$~s$-2$
%%% every two weeks. 141 days after the peak of the flare the very rapid
%%% changes in $\dot{\nu}$ started. From 141 days to 196 days after the flare,
%%% $\dot{\nu}$ dropped to $-$26.7$\times10^{-13}$~s$-2$ with an average drop of
%%% 5.6$\times10^{-13}$~s$-2$ every two weeks. $\dot{\nu}$ then started to rise
%%% again, and it fluctuated between $-$26$\times10^{-13}$~s$-2$ and
%%% $-$6$\times10^{-13}$~s$-2$ 4 times in the space of 450 days. During this
%%% period of unusual activity, there are four significant upward jumps in the
%%% spin-down. The fact that none of the four measured peak values of
%%% $\dot{\nu}$ are positive does not exclude the possibility of any of them
%%% indicating a glitch.
More specifically, we note that $\dot{\nu}$ was stable from 29 days
preceding the peak of the second flare, until 41 days following the flare,
fluctuating around the value $\sim$$-$2.3$\times10^{-13}$~s$^{-2}$ with
variations on the order of 0.08$\times10^{-13}$~s$^{-2}$ every two weeks.
Then, from the 41 days to 141 days after the flare, $\dot{\nu}$ dropped an
average of 2$\times10^{-13}$~s$^{-2}$ every two weeks. The very rapid
changes in $\dot{\nu}$ started 141 days after the peak of the flare. From
141 to 196 days after the flare, $\dot{\nu}$ dropped an average of
5.6$\times10^{-13}$~s$^{-2}$ every two weeks. $\dot{\nu}$ then fluctuated
between $-$26$\times10^{-13}$~s$^{-2}$ and $-$6$\times10^{-13}$~s$^{-2}$ 4 times
in the space of 450 days. During this period of unusual activity, there are
four significant upward jumps in $\dot{\nu}$. Although none of the measured
peak values of $\dot{\nu}$ is positive, spin-up glitches could still have
occurred between measurements.

%% 4th panel para 3 %%
This panel also shows that the frequency derivative stabilized between 2004
and 2007. It then appears to have decreased before the large glitch
associated with the third flare. However, this decrease stops in the two
weeks preceding the flare (see Section~\ref{sec:timf3}).

%% NUDOT NOISE %%
After the glitch, $\dot{\nu}$ increased by
$\sim$0.33$\times10^{-13}$~s$^{-2}$ every week, rising from
$\sim$$-$7.7$\times10^{-13}$~s$^{-2}$ to $-$2.9$\times10^{-13}$~s$^{-2}$ in
$\sim$130 days before starting to fall continuously again at the same rate.
A preliminary analysis of the most-recent data shows that in 2008 May, the
pulsar appears to have entered a new noisy phase (not shown in the Figure).
Weekly variations in $\dot{\nu}$ starting roughly a year after the onset of
the third flare are similar to, but a factor of $\sim$2 smaller, than the
variations observed starting 141~days after the peak of the second flare.
This noisy phase was still ongoing as of 2008 November~17. 
%% the date of submission of this paper.
%%% 350 days after the onset of the flare, $\dot{\nu}$ started dropping
%%% faster, with an average weekly drop of 0.6$\times10^{-13}$~s$^{-2}$ until it
%%% reached $-$15.5$\times10^{-13}$~s$^{-2}$. Then, 395~days after the onset of
%%% the flare, $\dot{\nu}$ rose with an average weekly variation of
%%% 1.4$\times10^{-13}$~s$^{-2}$ for a few weeks, before it started dropping
%%% again at the same rate. The weekly variations starting 395~days after the
%%% onset of the third flare were a factor of $\sim$2 smaller than the ones
%%% observed starting 141~days after the peak of the second flare. They also
%%% never reached values as low as $-$26$\times10^{-13}$~s$^{-2}$. However the
%%% noisy phase was still ongoing as of mid-2008, date of completion of this
%%% paper.

\subsection{Timing Around the First Flare}
\label{sec:timf1}

\begin{figure*}
\begin{center}
\includegraphics[scale=0.6]{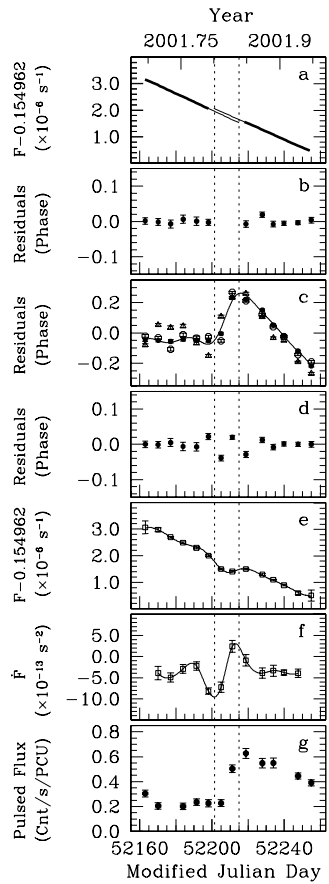}
\caption {Timing properties of AXP 1E~1048.1$-$5937 near the onset of the
first flare (MJD 52254 $-$ 52163). The two dotted lines enclose the data
from the week preceding and following the onset of the flare.
({\emph{a}}) Frequency versus time, showing the \texttt{TEMPO}-obtained
pre-flare and post-flare ephemerides, not including the data points between
the dotted lines. The lines shown between the double dotted lines are
extentions of the pre-flare and the post-flare ephemerides. 
%%% The corresponding residuals are shown in {\bf{panel~b}}.
({\emph{b}}) Timing residuals corresponding to panel~a.
%%%
({\emph{c}}) Data
points: timing residuals for three different sets of TOAs obtained after
subtracting the TOAs from the pre-flare ephemeris (see
Section~\ref{sec:timf1} for details). Solid curve: the spline that best fit
the pre-flare and post-flare TOAs subtracted from the pre-flare ephemeris.
({\emph{d}}) Timing residuals obtained after subtracting the original
TOAs from the spline. ({\emph{e}}) Frequency
obtained by evaluating the derivative of the spline shown in panel c.
({\emph{f}}) Frequency derivative obtained by
evaluating the second derivative of the spline shown in panel c.
({\emph{g}}) The 2$-$10~keV RMS pulsed flux of the pulsar near the onset
of the first flare. 
\label{plot-flare1timing}}
\end{center}
\end{figure*}

In this Section we describe the analysis of the TOAs in the 14 weeks
surrounding the onset of the first flare (MJD 52254 $-$ 52163). We show here
that a previously unreported and puzzling timing anomaly occured and was
coincident with the rise of the flare. The results are presented in
Figure~\ref{plot-flare1timing}.

%%% [[LINEBREAK ASTROPH]].
%%% delnu/nu sim 6.5196483441619947e-07 delnu sim 1.010309651494623e-07
In panel a, two lines, representing two \linebreak
ephemerides, are plotted. The left
ephemeris is obtained by fitting a frequency and a frequency derivative
through the pre-flare data, excluding the data between the dotted lines. The
right ephemeris is obtained by fitting the same parameters through the
post-flare data, again excluding the data between the dotted lines. If we
extend both ephemerides toward each other, it appears that a spin-up glitch
of size $\Delta\nu \sim$~1~$\times$~10$^{-7}$~s$^{-1}$ occured near the
onset of the flare. The residuals obtained from the two fits are presented
in panel~b.

In Section~\ref{sec:profiles}, we show that the pulsar underwent pulse
profile changes near the onset of the flares. Because we could not be
certain that our TOAs, obtained by cross-correlating the profiles of the
individual observations with a long-term template, were not affected by
pulse profile changes, we created two additional sets of TOAs. The first
additional set was obtained by aligning the tallest peak in the each profile
with the tallest peak in the template, and extracting a phase offset 
%%% which is in turn converted to a pulse arrival time. 
The second additional set was obtained by aligning the lowest point in each
profile with the lowest point in the template.
%%% and extracting a TOA.

In panel c, we subtracted all three sets of TOAs from the pre-flare
ephemeris and plotted the residuals. The points marked with solid
circles represent the residuals obtained from the original set of
TOAs. The points marked with empty circles and empty triangles
represent the residuals obtained from the two additional sets of TOAs.
While the scatter in the residuals corresponding to the additional
sets of TOAs is large, note how all three sets of residuals follow the
same trend, indicating that it is unlikely to be caused by pulse
profile changes.  However, it cannot be ruled out that the trend is
caused by the motion of the active region. The difference in phase
between each solid circle and the corresponding empty circle and empty
triangle represents our uncertainty in determining a fiducial point on
the pulsar. Also note that subtracting a full phase turn from all
post-flare residuals, which would yield a different timing solution,
would require a non-zero phase jump to have occured near the onset of
the flare, which would imply an unphysically large torque on the star.

Assuming the pulse numbers on which the residuals in panel c are based are
correct, we fit the pulse arrival times from the 14 weeks surrounding the
start of the flare with a spline. The spline subtracted from the pre-flare
ephemeris is the curve shown in panel c. The residuals after subtracting the
TOAs from the spline are shown in panel d. These residuals are clearly not
featureless.

The first derivative of the obtained spline, which is the frequency of the
pulsar, is shown in panel e. Notice the anomalous ``dip'' in frequency
surrounding the onset of the flare. The rapidly changing frequency
derivative is shown in panel f. The rms pulsed flux is shown for reference
in panel g (see Section~\ref{sec:flux} for more details on how the pulsed
flux is calculated). Notice how the dip in the frequency of the pulsar
started before the rise in the pulsed flux.
%%% date of the last observation with ok flux (slightly anomalous toa) is
%%% 52205. Date of the next observation (high flux, definitely off timing wise)
%%% is 52211. These are surrounded by the dotted lines.

%%% To summarize, when the ephemeris that fits the data during the decay of the
%%% first pulsed flux flare is extended backward to the time before the flare,
%%% and when the data collected during the rise and just before the rise of the
%%% flare are ignored, it looks as though a glitch occurred. Including these data
%%% however shows that just before and during the rise of the flare, the
%%% frequency was {\emph{smaller}} than that predicted by the pre-flare
%%% ephemeris. The frequency then rose and stabilized into the post-flare
%%% ephemeris.
To summarize, a timing anomaly occured near the onset of the
first flare. Careful analysis shows that it is not consistent with a simple
spin-up glitch, but with a gradual slow down lasting 2$-$3 weeks, followed by
a recovery. The rotational event appears to have preceded the flux event.

\subsection{Timing Around the Second Flare}
\label{sec:timf2}

\begin{figure*}
\begin{center}
\includegraphics[scale=0.8]{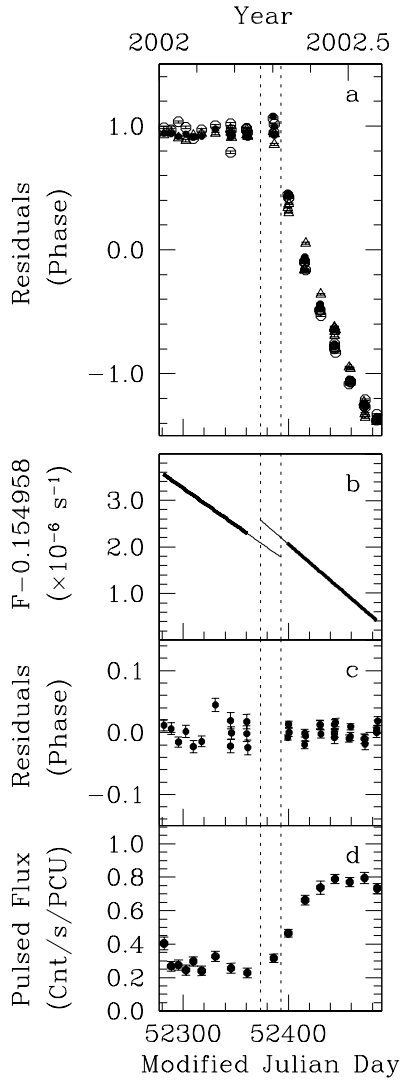}
\caption {Timing properties of AXP 1E~1048.1$-$5937 near the onset of the
second flare (MJD 52282 $-$ 52485). The two dotted lines enclose the
data from the week when the pulsed flux started rising. ({\emph{a}})
Timing residuals for three different sets of TOAs obtained after subtracting
the TOAs from the pre-flare ephemeris (see Section~\ref{sec:timf2} for details).
({\emph{b}}) Frequency versus time obtained from the original set of TOAs showing the
\texttt{TEMPO}-obtained pre-flare and post-flare ephemerides, not including the data
points between the dotted lines.
%%% The corresponding residuals are shown in {\bf{panel~c}}.
({\emph{c}}) Timing residuals corresponding to panel~b.
%%%
({\emph{d}}) The 2$-$10~keV RMS pulsed flux of the pulsar near the
onset of the second flare.
\label{plot-flare2timing}}
\end{center}
\end{figure*}

In this Section we describe our analysis of the TOAs in the 28 weeks
surrounding the onset of the second flare (MJD 52282 $-$ 52485). We have
discovered that a likely spin-up glitch occured during the week when the
pulsed flux started rising. We found this glitch while we were trying to fit
all available data with short simple overlapping ephemerides and encountered
a discontinuity. The results are presented in
Figure~\ref{plot-flare2timing}.

Once again, because of pulse profile changes around the start of the
flare, we generated two additional sets of TOAs by correlating the highest
and lowest points of the individual pulse profiles with the long-term
template and extracting phase differences. The residuals after subtracting
all three sets of TOAs from the pre-flare ephemeris are shown in panel a of
Figure~\ref{plot-flare2timing}. Once again, the scatter in the residuals
obtained from the additional sets of TOAs is larger than that obtained from
the standard TOAs, but all three sets follow the same trend.

The trend in the residuals shown in panel a indicates that a glitch occured.
However, due to the finite resolution of the data (sets of three
closely-spaced observations obtained every two weeks, starting in 2002
March), which is particularly problematic given the extreme timing noise
of this source, a rapid non-instantaneous variation cannot be ruled out. The
curvature following the glitch is due to a change in the frequency
derivative rather than glitch recovery. Because the largest pulse profile
changes occured in the week the pulsed flux started rising (see
Section~\ref{sec:flux}), there is large scatter in the three standard TOAs
obtained then. Because of this scatter, it was not possible to determine if
the glitch occured before or after the pulsed flux started to rise. The
glitch epoch was MJD 52386.0~$\pm$~1.5. The dates of the first three
observations having a larger pulsed flux than the pre-flare long-term
average are MJD 52385.5, 52386.6, and 52386.7. The change in the frequency
at the time of the glitch was
$\Delta\nu$~=~4.51(14)~$\times$~10$^{-7}$~s$^{-1}$
($\Delta\nu/\nu$~=~2.91(9)~$\times$~10$^{-6}$). The change in frequency
derivative was $\Delta\dot{\nu}$~=~$-$4.10(15)~$\times$~10$^{-14}$~s$^{-2}$.

The pre-flare and post-flare ephemerides are shown in panel b of
Figure~\ref{plot-flare2timing}. Note the
difference in slope between them. The residuals are shown in
panel~c. The pulsed flux is shown for reference in panel~d. Each pulsed
flux data point is the average of the pulsed flux values obtained from three
closely spaced onservations.

To summarize, a glitch, or a very rapid change in the frequency, as well as
a significant change in the frequency derivative, occured during the week
the second pulsed flux flare started rising. Because of the large
uncertainty on the glitch epoch, which is due to the pulse profile changes
near the onset of the flare, it is not possible to determine which happened
first, the rise in the pulsed flux, or the frequency jump.

\subsection{Timing Around the Third Flare}
\label{sec:timf3}

\begin{figure*}
\begin{center}
\includegraphics[scale=0.6]{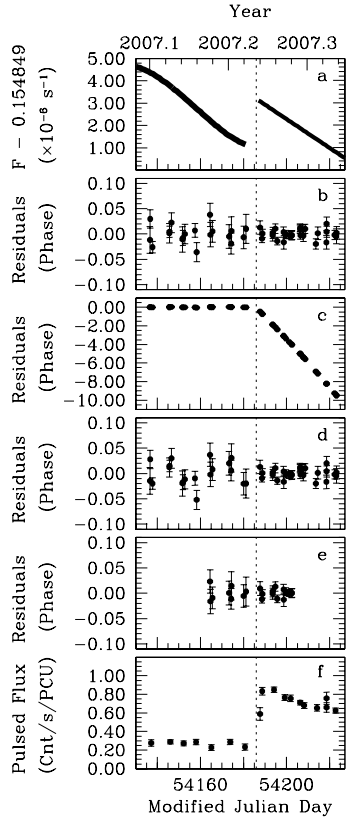}
\caption {Timing properties of AXP 1E~1048.1$-$5937 near the onset of the
third flare (MJD 54131 $-$ 54223). The dotted line marks the epoch of a
large glitch (see Section~\ref{sec:timf3} for details). ({\emph{a}})
Frequency versus time of the pre-flare and post-flare ephemerides. The
pre-flare ephemeris consists of a frequency and three frequency
derivatives. Notice how the curve is flatter in the two weeks preceding the
flare. The post-flare ephemeris consists of a frequency and a single
frequency derivative. 
%%% The corresponding timing residuals are shown in {\bf{panel~b}}. 
({\emph{b}}) Timing residuals corresponding to panel~a.
%%%
({\emph{c}}) Residuals obtained after subtracting
pre-flare and post-flare TOAs from a 
pre-flare ephemeris consisting of a frequency and frequency derivative.
The change in the slope marks the
occurence of the glitch. ({\emph{d}}) Timing residuals obtained after
fitting a glitch through the data for the 14 weeks surrounding the glitch.
The RMS phase residual is 1.6\%. ({\emph{e}}) Timing residuals obtained
after fitting a glitch through the data for the 6 weeks surrounding the
glitch. The RMS phase residual is 0.98\%. ({\emph{f}}) The 2$-$10~keV RMS pulsed
flux of the pulsar near the onset of the third flare (see
Section~\ref{sec:timf3} for details).
\label{plot-glitch}}
\end{center}
\end{figure*}

In this Section we report on our analysis of the TOAs in the 14 weeks
surrounding the onset of the third flare (MJD 54131~$-$~54223). We show that
a large spin-up glitch occured coincident with the rise of the pulsed flux.
The results are presented in Figure~\ref{plot-glitch}.

We first plotted the pre-flare and post-flare ephemerides in panel~a. The
pre-flare ephemeris consists of a frequency and three frequency derivatives.
The post-flare ephemeris consists of a frequency and a single frequency
derivative. The residuals are shown in panel~b. Note how in panel~a, the
pre-flare curve appears to flatten in the two weeks preceding the glitch,
indicating that the frequency derivative was becoming less negative. This
argues that it is important to choose data as close to the glitch as
possible when fitting for the glitch parameters.

In panel c, we show the pre-glitch and the post-glitch timing residuals
after subtracting the TOAs from an ephemeris that includes the frequency and
frequency derivative that best fit the pre-glitch data. The observed trend
in the residuals clearly indicates that a large glitch occured. To obtain
the glitch parameters, we performed two different fits with \texttt{TEMPO}.

For the first fit, we included data from the 14 weeks surrounding the glitch
epoch. We subtracted the TOAs from an ephemeris consisting of the best-fit
$\nu$, $\dot{\nu}$, and discrete jump in $\nu$ and $\dot{\nu}$ at the glitch
epoch. The timing residuals for the first fit are shown in panel d. For the
second fit, we included data from the 6 weeks surrounding the glitch epoch.
We subtracted the TOAs from an ephemeris consisting of the same set of
parameters. The timing residuals for the second fit are shown in panel e. As
expected, the best-fit jump in $\dot{\nu}$ at the glitch epoch was
significantly larger for the first fit than for the second fit
(1.76(8)$\times$10$^{-13}$~s$^{-2}$ versus 6(4)$\times$10$^{-14}$~s$^{-2}$).
This is because of the rapid change in the frequency derivative in the few
weeks preceding the glitch.

%%% [[LINEBREAK ASTROPH]].
%%% [[LINEBREAL APJ]]
From the second fit, the total frequency jump observed at the glitch epoch
was \linebreak $\Delta\nu$~=~2.52(3)~$\times$~10$^{-6}$~s$^{-1}$
($\Delta\nu/\nu$~=~1.63(2)~$\times$~10$^{-5}$)\footnote{This is different
from the value in \cite{dkgw07UPDATED} because of a typographical error: the
authors reported the value of $\Delta\nu$ instead of reporting the value of
$\Delta\nu/\nu$.}. The glitch epoch, determined by setting the phase jump
to zero at the time of the frequency jump, is MJD 54185.912956 (2007
March~26). For a complete list of the fit parameters, see
Table~\ref{tableglitch}.

The pulsed flux is shown for reference in panel~f of
Figure~\ref{plot-glitch}. Each plotted pulsed flux data point is the average
of the pulsed flux values obtained from three closely spaced onservations,
except in two instances (see Section~\ref{sec:flux}). The date of the last
pre-flare observation was MJD 54181.32. The date of the first observation
with a large pulsed flux is MJD 54187.67. As explained in
Section~\ref{sec:flux}, it is difficult to determine if the pulsed flux of
the latter observation is lower than the pulsed flux peak, due to noise. 
%%% The glitch epoch was determined to be 54185.912956. 
Once again, we cannot determine whether the glitch occured before or after
the pulsed flux started rising.

To summarize, a large glitch occured on MJD 54185, two days before the first
observation having a large pulsed flux. The change in the frequency
derivative at the time of the glitch was not significant, but it was
preceded by three weeks where the magnitude of $\dot{\nu}$ was decreasing,
which followed a rapid decrease that lasted several weeks. Because of the
possibility that the pulsed flux of the first observation after the onset of
the flare is consistent with the peak of the flare, we were not able to
determine which happened first, the rise in the pulsed flux, or the glitch.

%%% \item parameters long fit
%%% deltanu/nu = 1.707(9)e-05
%%% delta nudot 1.75612981E-13  1      7.90514625E-15
%%% \item parameters short fit
%%% delta nu/nu = 1.63(2)e-5
%%% delta nudot 6.29419427E-14  1      3.84542565E-14 <<<
%%% epoch 54185.912956
%%% \item REFER TO GLITCH TABLE FOR ADDITIONAL PARAMETERS
%%% F0      0.1548496877576173  (err 0.0000000614599808)
%%% F1     -8.234460077895E-13  (err 5.151610402117E-14)
%%% PEPOCH        54185.912956
%%% START            54164.545
%%% FINISH           54202.475
%%% GLEP_1      54185.912956
%%% GLPH_1            0.000000  1            0.011454
%%% GLF0_1      2.52838230E-06  1      3.13280767E-08
%%% GLF1_1      6.29419427E-14  1      3.84542565E-14 <<<
%%% NTOA                    21

\section{Pulse Profile Study: Analysis and Results}
\label{sec:profiles}

\cite{tgd+08} reported pulse profile changes in 1E~1048.1$-$5937 from
imaging data near the third flare. In this Section we confirm their findings
and report on additional pulse profile changes near the first two flares.

We performed a first pulse profile analysis using FTOOLS version
5.3.1\footnote{http://heasarc.gsfc.nasa.gov/ftools}. Data from PCU~0 were
included in the analysis up to 2000 May 12, when it lost its propane layer.
Data from PCU~1 were included in the analysis up to 2006 December 25, when
it lost its propane layer. We used the procedure described in detail in
\cite{dkg07UPDATED} to extract a pulse profile for each observation in the
2$-$10~keV band. We used 64 phase bins.
%%% in countrate per pcu.
When a local ephemeris was not available, we folded the data at a pulsar
period extracted from a periodogram.  We verified that the results of the
folding are not very sensitive to the precise period used. We then aligned
the 64-bin profiles with a high signal-to-noise template using a
cross-correlation procedure similar to that described in
Section~\ref{sec:timf0}.

\begin{figure*}
\begin{center}
\includegraphics[scale=0.75]{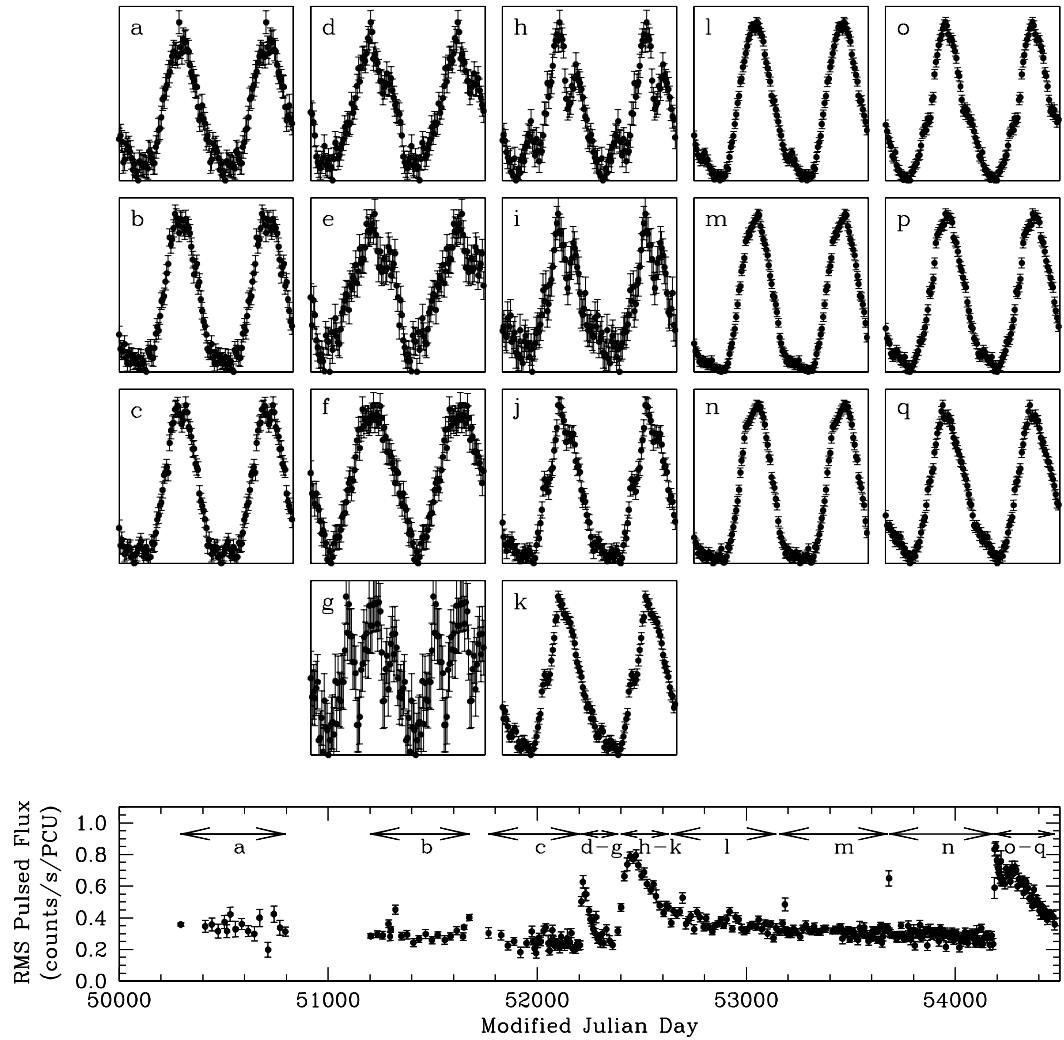}
\caption {Normalized 2$-$10~keV pulse profiles of 1E~1048.1$-$5937 from 1997
to 2008. The letter shown in the top-left corner of each plot refers to the
time segments marked by arrows in the bottom plot, where
the 2$-$10~keV RMS pulsed flux is shown for reference.
\label{plot-profiles}}
\end{center}
\end{figure*}

1E~1048.1$-$5937 was monitored with {\emph{RXTE}} from 1997 to 2008. To do
the first pulse profile analysis, we divided this time span into many
segments, shown with letters at the bottom of Figure~\ref{plot-profiles}.
%%% We divided the first three years of monitoring into segments~a, b, and~c.
%%% Segments~d and~e covered the rise and the fall of flare~1. Segment~f covered
%%% the time between flares~1 and~2. Segment~g covered the week where the pulsed
%%% flux started to rise again. Segments~h, i, j, and~k covered the rise, the
%%% two halves of the peak, and the fall of flare~2. Segments~l, m, and~n
%%% covered the three years of quiescence. Finally, segments~o, p, and~q covered
%%% the recovery from the third flare. 

For each time interval, we summed the aligned profiles, subtracted the DC
component from the summed profile, and scaled the resulting profile so that
the value of the highest bin is unity and the lowest point is zero. The
results are presented in Figure~\ref{plot-profiles} with the time intervals
marked in the top left corner of each profile. The different profile
qualities are due to the segments having different total exposure, and to
changes in the pulsed flux of the pulsar.
%%% D92006-02-54-00 rise obs 03/28/2007
%%% D92005-02-05-00 04/09day/2007 too observation, 6ks long, 2 pcus all time.
%%% D92006-02-59-02 05/03/2007 2ks long 3 PCUS operational whole duration

\begin{figure*}
\begin{center}
\includegraphics[scale=0.9]{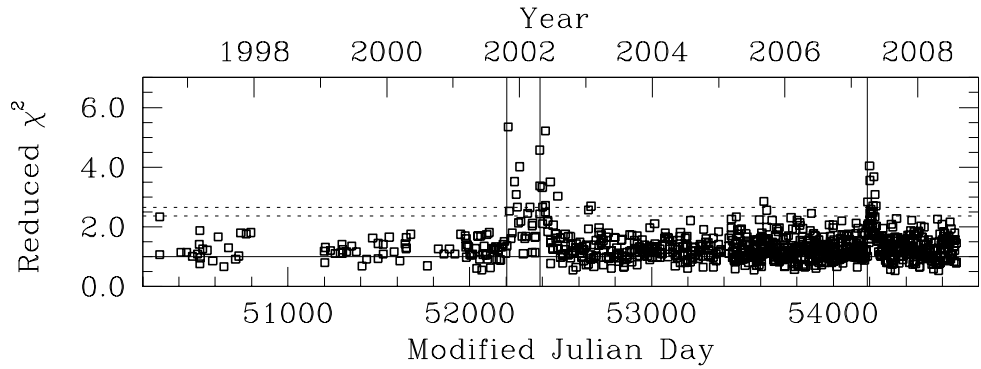}
\caption{Reduced ${\chi}^{2}$ statistics versus time, calculated after
subtracting the scaled and aligned profiles of the individual observations
from a high signal-to-noise template. The solid vertical lines indicate the
onsets of the flares. The solid horizontal line indicates a reduced
${\chi}^{2}$ of 1. The lower dotted line corresponds to the 2~$\sigma$
significance level. 
%%% Any point on this line indicates that the difference
%%% between the pulse profile of the corresponding observation and the long-term
%%% average profile is significant on the 2~$\sigma$ level. 
The upper dotted
line corresponds to the 3~$\sigma$ significance level.
\label{plot-chi}}
\end{center}
\end{figure*}

To look for pulse profile changes on a smaller timescale, we performed a
second pulse profile analysis. We extracted a pulse profile for each
observation in the 2$-$10~keV band using all available PCUs to maximize the
signal to noise. We used 32 phase bins. We aligned the obtained profiles
with the high signal-to-noise template and subtracted the respective average
from each of the aligned profiles and from the template. For each
observation, we then found the scaling factor that minimized the reduced
${\chi}^2$ of the difference between the scaled profile and the template.
The obtained reduced ${\chi}^2$ values are plotted in Figure~\ref{plot-chi}.

Figure~\ref{plot-profiles} shows for the first time that the broad pulse
profile of 1E~1048.1$-$5937 developed a small side-peak during the rise and
the fall of the first flare (segments~d and~e). The rise of the second flare
(segments~g and~h) was marked by large profile changes in which the pulse
profile was clearly multipeaked. The pulse profile slowly returned to its
long-term average shape while the flare was decaying (segments~i, j, and~k).
There were no significant pulse profile changes in the following three years
of quiescence (segments~l, m, and~n), although the profiles in segments~m
and~n seem to have triangular peaks, more so than in segments~b, ~c, and~l.
%%% We divided these three years into shorter time intervals and repeated the
%%% same analysis. We discovered that low-level short-term pulse profile
%%% variations occured in segment~l, around the time of the large torque
%%% variations seen in Figure~\ref{plot-timing}.
%%% There were no significant short-term pulse profile variations in segments~m
%%% and~n.
Figure~\ref{plot-chi} confirms the pulse profile changes near the first two
flares, and additionally suggests that small occasional profile changes may
occur in individual observations throughout segments~a, b, c, l, m, and~n,
but only at the $\sim$2$-$3$\sigma$ level.

\begin{figure*}
\begin{center}
\includegraphics[scale=0.99]{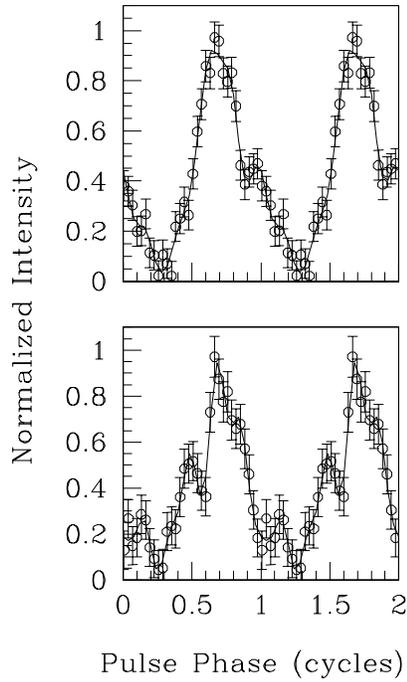}
\caption {Normalized 2$-$10~keV pulse profiles from two observations taken
during the decay of the third flare. The first observation was taken on 2007
April~09 (14 days after the glitch epoch). The second observation was taken
on 2007 May~03 (38 days after the glitch epoch). The multiple peaks in the
profile are obvious. \label{plot-twoprofiles}}
\end{center}
\end{figure*}

It appears from Figure~\ref{plot-profiles} that the pulse profiles in
segments~o, p, and~q, corresponding to the decay of the third flare, were
stable and presented no significant deviations from the long-term average on
long timescales. However, Figure~\ref{plot-chi} shows that many significant
pulse profile changes occured on short timescales during the decay of the
third flare. The changes are clearly visible in the pulse profiles of
individual observations having a high signal-to-noise ratio (particularly
long observations, or observations with a large numbers of operational
PCUs). An example of two such profiles is presented in
Figure~\ref{plot-twoprofiles}. The top profile is obtained from a 6~ks-long
observation taken on 2007 April~09 (14 days after the glitch epoch) with two
operational PCUs. The second profile is obtained from a 2~ks-long
observation taken on 2007 May~03 (38 days after the glitch epoch). These
short-term pulse profile changes are similar to the ones reported in
\cite{tgd+08}. They were seen mostly in the first two months following the
onset of the flare, and occured less often in the next months, although this
may be partially due to the reduction in signal-to-noise ratio due to the pulsed
flux falling. Several months after the flare small occasional profile
changes may be present, but only at the $\sim$2$\sigma$ level.

\section{Pulsed Flux Study: Analysis and Results}
\label{sec:flux}

%%% [[LINEBREAK ASTROPH]].
To obtain a pulsed flux time series for \linebreak
1E~1048.1$-$5937,
%%% We used FTOOLS version 5.3.1\footnote{http://heasarc.gsfc.nasa.gov/ftools}.
%%% to create a barycentered time series in the 2$-$10~keV band for each
%%% observation. Data from PCUs 0 and 1 were excluded after the loss of their
%%% respective propane layers for which an independent analysis of AXP 4U 0142+61
%%% revealed spectral modeling irregularities \citep{dkg07UPDATED}.
for each observation, we created a pulse profile (in units of count rate per
PCU) using the same procedure as in Section~\ref{sec:profiles}.  Data from
PCUs 0 and 1 were excluded after the loss of their respective propane
layers, because an independent analysis of data from PCU~0 of AXP 4U~0142+61
revealed spectral modeling irregularities after the loss of the propane
layer \citep{dkg07UPDATED}. Pulse profiles were generated in three bands:
2$-$4~keV, 4$-$10~keV, and 2$-$10~keV. For each folded profile, we
calculated the RMS pulsed flux,

\begin{equation}
F_{RMS} = 
{\sqrt{2 {\sum_{k=1}^{n}}
(({a_k}^2+{b_k}^2)-({\sigma_{a_k}}^2+{\sigma_{b_k}}^2))}},
\label{eq:f1}
\end{equation}
\noindent where $a_k$ is the $k^{\textrm{\small{th}}}$ even Fourier
component defined as $a_k$ = $\frac{1}{N} {\sum_{i=1}^{N}} {p_i} \cos {(2\pi
k i/N})$, ${\sigma_{a_k}}^2$ is the variance of $a_k$, $b_k$ is the odd
$k^{\textrm{\small{th}}}$ Fourier component defined as $b_k$ = $\frac{1}{N}
{\sum_{i=1}^{N}} {p_i} \sin {(2\pi k i/N})$, ${\sigma_{b_k}}^2$ is the
variance of $b_k$, $i$ refers to the phase bin, $N$ is the total number of
phase bins (here $N$=64), $p_i$ is the count rate in the $i^{\textrm{\small{th}}}$ phase
bin of the pulse profile, and $n$ is the maximum number of Fourier harmonics
used; here $n$=5.

%%% The second method yields an estimate of the ``area'' pulsed flux.
%We verified using an area-based pulsed flux estimator (Archibald
%et al. in prep.) that the trends seen in the RMS pulsed flux of
%1E~1048.1$-$5937 are not a consequence of changes in the pulse profile. The
%results of the pulsed flux analysis are presented in Figure~\ref{plot-flux}.

\begin{figure*}
\begin{center}
\includegraphics[scale=0.75]{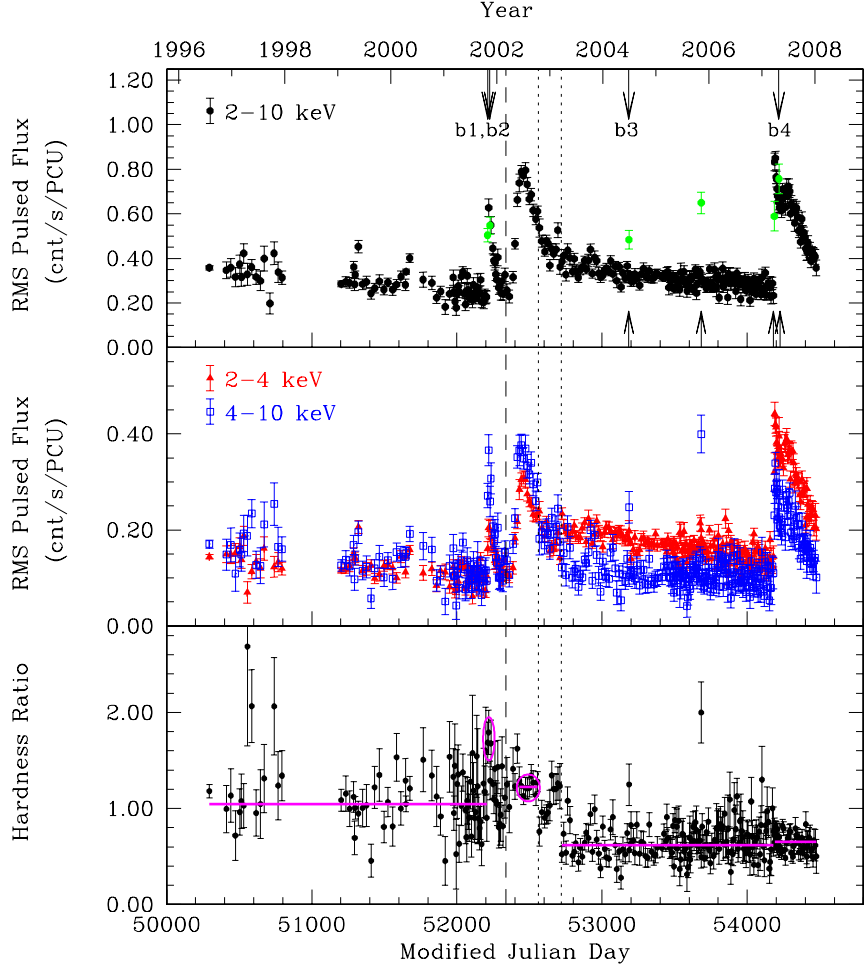}
\caption
{{\emph{Top:}} Pulsed flux time series in the 2$-$10~keV band. For
observations taken after 2002 March~02, we plotted the average of the pulsed
flux values of each three closely spaced observations, with the 4 exceptions
indicated by arrows along the bottom of the panel. All observations
containing bursts are indicated by arrows along the top of the panel. All
points indicated with an arrow are also coloured in green. {\emph{Middle:}}
Pulsed flux time series in the 2$-$4~keV band (red triangles) and in the
4$-$10~keV band (blue squares). {\emph{Bottom:}} Hardness ratio computed
from the pulsed flux in the energy range (4$-$10~keV)/(2$-$4~keV).
The hardness ratios near the peaks of the first two flares are 
marked with two magenta circles. 
%%% The weighted average hardness ratio for the years preceding the first flare
%%% is marked with a magenta horizontal line. The hardness ratios near the peaks
%%% of the first two flares are marked with two magenta circles. The hardness
%%% ratio for the 4 years preceding the third flare is also marked with a
%%% magenta horizontal line.  Finally, the hardness ratio after the onset of the
%%% third flare is marked with another magenta horizontal line.
{\emph{All panels:}} The dashed line indicates the epoch when we started
observing the source with sets of three closely spaced observations. The left
dotted line marks the location when the hardness ratio dropped, 342 days after
the peak of the second flare. The right dotted line marks the location when
the hardness ratio drops further, 500 days after the peak of the second
flare.
\label{plot-flux}}
\end{center}
\end{figure*}

We verified using an independent pulsed flux estimator calculated
from the area under the pulse (i.e. insensitive by definition to
> pulse shape) that the trends seen in the RMS pulsed flux of
1E~1048.1$-$5937 are not a consequence of changes in the pulse
profile. The results of the pulsed flux analysis are presented in
Figure~\ref{plot-flux}.

%%% \begin{equation}
%%% F_{AREA} =
%%% {a_0 - p_{\textrm{\small{min}}}}
%%% \label{eq:f2}
%%% \end{equation}
%%%
%%% where $a_0 = \frac{1}{N}{\sum_{i=1}^{N}} {p_i}$, $i$ refers to the phase
%%% bin, $N$ is the total number of phase bins, $p_i$ is the count rate in the
%%% $i^{\textrm{\small{th}}}$ phase bin of the pulse profile, and
%%% $p_{\textrm{\small{min}}}$ is the average count rate in the off-pulse
%%% interval of the profile, determined by cross-correlating with a high
%%% signal-to-noise template, and calculated in the Fourier domain after
%%% truncating the Fourier series to 5 harmonics.

%%% We used techniques described in Archibald et al. (in prep) to evaluate both
%%% quantities for each band in each observation. The two pulsed flux time
%%% series differed only by a constant scaling factor. So we chose to report the
%%% RMS pulsed flux because it has smaller uncertainties. The results are
%%% presented in Figure~\ref{plot-flux}.

In the top panel, we show the pulsed flux results in the 2$-$10~keV band.
For observations taken before 2002 March~02 (date marked with a dashed
line), we plotted the pulsed flux values obtained from individual
observations. After 2002 March~02, we plotted the average of the pulsed flux
values of each set of three closely spaced observations, with the exception of 4
observations. The 4 observations are indicated with arrows located along the
bottom of the panel. The first observation, on 2004 June~29, was not
averaged with its neighbors because a burst occured within the observation
\citep{gkw06}. The second, on 2005 November~08, is an observation with an
anomalously high pulsed flux. The third, on 2007 March~28, is the first
observation that is part of the most recent pulsed flux flare. The fourth
observation, on 2007 April~28, also contained a burst. In each of these
cases, we have singled out the abnormal observation, and averaged the other
two that were part of the same set. Each of these exceptions is discussed
below and in Section~\ref{sec:bursts}. 
%%% All observations containing bursts, are indicated by arrows along the top of
%%% the panel \citep{gk04,gkw06}. All points indicated with an arrow are also
%%% coloured in green.

Also in the top panel, the pulsed flux time series obviously has significant
structure. The most obvious features are the three long-lived flares.
\cite{gk04} estimated the peak flux of the first flare to occur at MJD
52218.8~$\pm$~4.5, with a risetime of 20.8~$\pm$~4.5 days and a fall time of
98.9~$\pm$~4.5 days. They estimated the peak flux of the second flare to
occur at MJD 52444.4~$\pm$~7.0, with a risetime of 58.3~$\pm$~7.0 days and a
fall time greater than 586 days. In fact, we can see in the top panel that
the second flare continued to decay slowly, and that the pulsed flux had not
returned to its pre-flares value by the time the third flare occured. 
However, the pulsed flux in the year prior to the flares was low compared to
the previous years, making it unclear what the real quiescent flux level
is.

Here we estimate the peak pulsed flux of the third flare to have occured at
MJD 54191.6~$\pm$~3.1, with an upper limit on the risetime of 7.3 days. The
three observations obtained in the last week before the flare all had a
pulsed flux consistent with quiescence. The three observations obtained 7
days later all had a significantly higher pulsed flux. The first of these
three observations, occured 1.75 days after the determined glitch epoch and
had a lower pulsed flux than the second observation. The two observations were
separated by 22 hours. It is possible that the first observation is part of
the rise of the flare, which would imply a resolved rise with a risetime significantly smaller
than 7 days. However the value of the pulsed flux for that observation
before the binning is less than 3$\sigma$ away from that of the
following observation, and the scatter in the unbinned data near the start
of the flare is large. We also estimate the fall time of that flare to be
greater than 288 days (date of the last observation included in this paper)
since the pulsed flux had not returned to its pre-flare value.

We estimate that the three flares had peak pulsed fluxes of 2.32~$\pm$~0.15,
2.90~$\pm$~0.07, and 3.13~$\pm$~0.10 times the quiescent pulsed flux for the
2$-$10~keV band. By ``quiescent pulsed flux'' we mean the average pulsed
flux from the year preceding the first flare and from the year preceding the
third flare.

An anti-correlation between the total flux and the pulsed fraction has been
reported for this source \citep{tmt+05}. \cite{tgd+08} used imaging
observations from 1E~1048.1$-$5937 to derive the following anti-correlation
in the 2$-$10~keV band:

\begin{equation}
F_{tot} = A \times (F_{RMS}/a)^{1/(1+b)},
\label{eq:eq4}
\end{equation}

\noindent where $F_{tot}$ is the total flux of the source in erg/s/cm$^2$,
$F_{RMS}$ is the RMS pulsed flux in counts/s/PCU, $a$ and $b$ are constants
($a$=1.53, $b$=$-$0.46),
and $A$ is a constant scaling factor ($A$$\sim$125). Note that there were no imaging
observations obtained in a data mode suitable for extracting pulsed fractions
from near the peaks of the first two flares; the parameters in
the above equation were obtained on the basis of the third flare only. This
information allows us to scale our pulsed fluxes to estimate the total
energy released in each flare, assuming that the relation for the third
flare holds for the first two as well.
%%% 5~kpc \citep{opk01}, 2.7~kpc \citep{gmo+05}, \citep{dv06ax} 9.0KPC AS WELL
Assuming a distance of 2.7~kpc \citep{gmo+05}, and assuming roughly linear
decays (see Section~\ref{sec:discussfluxes}), we find a total energy release
of $\sim$~4.4$\times$~10$^{40}$~erg for the first flare,
$\sim$~3.1$\times$~10$^{41}$~erg for the second, and
$\sim$~3.9$\times$~10$^{41}$~erg for the third, all in the 2$-$10~keV band.
For a distance of 9~kpc \citep{dv06ax}, these numbers become
$\sim$~4.8$\times$~10$^{41}$~erg, $\sim$~3.5$\times$~10$^{42}$~erg,
and~$\sim$~4.3$\times$~10$^{42}$~erg.
%%% >>> (70*86400)*(21-4.5)*0.5e-12*(4*pi*((5*3.08568025e21)**2))
%%% 1.4925106725612301e+41 (other paper 2.7e40)
%%% >>> (260*86400)*(36.5-4.5)*0.5e-12*(4*pi*((5*3.08568025e21)**2))
%%% 1.075124571057094e+42 (other paper 2.8e41)
%%% >>> (350*86400)*(34-4.5)*0.5e-12*(4*pi*((5*3.08568025e21)**2))
%%% 1.3342140860774633e+42 (Cindy paper estimate 7e39 to 8e40 distance depend).
%%%
%%% >>> (70*86400)*(21-4.5)*0.5e-12*(4*pi*((2.7*3.08568025e21)**2))
%%% 4.3521611211885484e+40
%%% >>> (70*86400)*(21-4.5)*0.5e-12*(4*pi*((9*3.08568025e21)**2))
%%% 4.8357345790983865e+41
%%%
%%% >>> (260*86400)*(36.5-4.5)*0.5e-12*(4*pi*((2.7*3.08568025e21)**2))
%%% 3.1350632492024869e+41
%%% >>> (260*86400)*(36.5-4.5)*0.5e-12*(4*pi*((9*3.08568025e21)**2))
%%% 3.4834036102249849e+42
%%%
%%% >>> (350*86400)*(34-4.5)*0.5e-12*(4*pi*((2.7*3.08568025e21)**2))
%%% 3.8905682750018843e+41
%%% >>> (350*86400)*(34-4.5)*0.5e-12*(4*pi*((9*3.08568025e21)**2))
%%% 4.3228536388909819e+42
%%%

In the middle panel of Figure~\ref{plot-flux}, we show the pulsed flux results in 2$-$4~keV (red
triangles) and in 4$-$10~keV (blue squares). Note that in the years between
flares~2 and~3 there are two data points with a significantly high pulsed
flux in the 4$-$10~keV band. The corresponding dates are 2004 June~29 (MJD
53185) and 2005 November~08 (MJD 53682). The observation corresponding to
the first point contains the third burst detected from this source.
\cite{gkw06} reported a pulsed flux increase immediately following the
burst, and a slow decay within the 2~ks-long observation (see
Section~\ref{sec:bursts}). We did not detect a burst within the observation
corresponding to the second point, and we found no evidence for a decay in
the pulsed flux within the 1.5~ks-long observation. See
Section~\ref{sec:discussbursts} for more discussion.

%%% [[LINEBREAK APJ]]
In the bottom panel, we show the ratio of the two pulsed flux time series \linebreak
$H\equiv$~(4$-$10~keV/2$-$4~keV). The weighted average hardness ratio for
the years preceding the first flare is marked with a magenta horizontal line
with $H$~=~1.04$\pm$0.02. The hardness ratios near the peaks of the first
two flares are marked with two magenta circles and have $H$~=~1.6$\pm$0.1
and $H$~=~1.22$\pm$0.03, respectively. The hardness ratio for the 4 years
preceding the third flare is also marked with a magenta horizontal line with
$H$~=~0.62$\pm$0.01. Finally, the hardness ratio after the onset of the
third flare is marked with another magenta horizontal line at
$H$~=~0.65$\pm$0.01. It is clear from the middle and bottom panels that the
pulsed emission from 1E~1048.1$-$5937 had a harder spectrum near the peaks
of the first two flares compared to the pulsed emission preceding the
flares. It is also clear that 342 days after the peak of the second flare
(first vertical dotted line), this ratio dropped. It dropped again 500 days
after the peak of the second flare (second vertical dotted line) to a value
smaller than the pre-flares value, a value that was maintained until the
onset of the third flare. We verified that these changes in the hardness
ratio do not coincide with the epochs of gain change of {\emph{RXTE}}, nor
do they coincide with the dates of loss of the propane layers of PCUs~0
and~1. We also verified that there are no similar changes at the same epochs
in the other monitored AXPs.

%%%% \subsection{Comparison of {\emph{RXTE}} and {\emph{CXO}}-observed Spectral Changes}
%%%% \label{sec:cxo}

Note that the hardness ratios reported above are obtained from the pulsed
flux of {\emph{RXTE}}. Any changes in the hardness ratio of the total flux,
as observed by an imaging instrument, might not necessarily be reflected in
the behavior of the pulsed hardness ratio if the pulsed and persistent
spectra are different. 
%
%\cite{tgd+08} reported a significant increase in the
%hardness ratio obtained from {\emph{Chandra X-ray Observatory
%}}({\emph{CXO}}) data at the onset of the third flare. 
%The hardness ratio
%was defined as ($S+H$)/($S$$-$$H$) where $S$ and $H$ are countrates in the
%1$-$3~keV and 3$-$10~keV bands, and it increased by $\sim$14\% at the onset
%of the flare. With {\emph{RXTE}}, we observe a marginal increase in
%the (4$-$10~keV)/(2$-$4~keV) pulsed hardness ratio when we compare the
%pre-flare and the post-flare data. 
%
%This shows that the pulsed spectrum is evolving differently from the
%total phase-averaged spectrum. Indeed we found after analysing 
%data from six {\emph{CXO}} observations
%that the dependence of the pulsed fraction on energy is changing
%with time.
%
\cite{tgd+08} reported an increase in the (3$-$10~keV)/(1$-$3~keV)
hardness ratio obtained from {\emph{Chandra X-ray Observatory
}}({\emph{CXO}}) data at the onset of the third flare.  With
{\emph{RXTE}}, we observe a marginal increase in the
(4$-$10~keV)/(2$-$4~keV) pulsed hardness ratio when we compare the
pre-flare and the post-flare data. This can be attributed to the
pulsed spectrum having a different evolution from the total
phase-averaged spectrum. Indeed we found after analysing data from six
{\emph{CXO}} observations that the dependence of the pulsed fraction
on energy is changing with time.

\section{A New Burst}
\label{sec:bursts}

%% intro %%
Searching for bursts is part of our regular AXP monitoring routine.  For
each observation of 1E~1048.1$-$5937, we generated 31.25~ms lightcurves
using all Xenon layers and events in the 2--20~keV band. These
lightcurves were searched for bursts using the algorithm
introduced in \citet{gkw02} and discussed further in \citet{gkw04}. Four
bursts have been detected from this source. 
%%% The bursts were significant in each active PCU. 
The first two bursts occured on 2001 October~29 and 2001 November~14
\citep{gkw02}. One of these bursts was coincident with the rise of the first
pulsed flux flare, and the other with its fall (see Figure~\ref{plot-flux}).
The third burst occured on 2004 June~29, 740~days following the peak of the
second pulsed flux flare \citep{gkw04}. Here, we report on the detection of
a fourth burst\footnote{The burst search routine also returned several
candidate bursts with a significance several orders of magnitude smaller
than those reported for the published bursts. We do not report on the
analysis of these putative bursts here.}, which occured on 2008 April~28, 27~days
after the peak of the third flare.

\subsection{Burst Properties}

\begin{figure*}
\begin{center}
\includegraphics[scale=0.9]{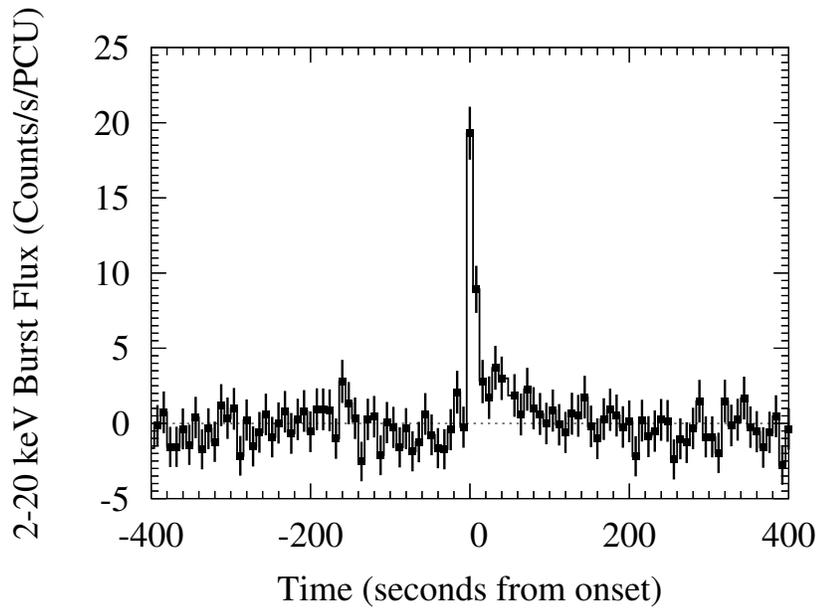}
\caption
{Background subtracted 2$-$20 keV burst light curve binned with 8~s time
resolution.
\label{plot-burstlc}}
\end{center}
\end{figure*}

%% time parameters %%
To analyse the burst, we created event lists in
FITS\footnote{\url{http://fits.gsfc.nasa.gov}} format using the standard
\texttt{FTOOLS}\footnote{\url{http://heasarc.gsfc.nasa.gov/docs/software/ftools/}}.
For consistency with previous analyses of SGR/AXP bursts, we extracted events
in the 2-20~keV band, and reduced them to the solar system barycenter.  We
subtracted the instrumental background using the model background lightcurve
generated by the \texttt{FTOOL} \texttt{pcabackest}. The model background
lightcurve generated by \texttt{pcabackest} only has 16-s time
resolution. We therefore fit the simulated background lightcurve to a fourth order
polynomial and subtracted this model from our high-time-resolution
lightcurves. Using the resulting light curve, we then subtracted an
additional background determined from a 300-s long interval ending 100-s
before the burst. The final background-subtracted burst lightcurve is shown
in Figure~\ref{plot-burstlc}. The burst temporal properties, namely peak time
($t_p$), peak flux ($f_p$), rise time ($t_r$), $T_{90}$, which is the time
from when 5\% to 95\% of the total burst counts have been collected, and
$T_{90}$ fluence were determined using the methods described in
\citet{gkw04}. The peak flux and $T_{90}$ fluence were then determined in
units of erg/s/cm$^2$ and erg/cm$^2$, respectively, assuming a
power-law spectrum (see below). The burst properties are listed in
Table~\ref{table:burst}. The burst risetime, 955~ms, calculated using a
linearly rising model, is longer by a factor of $\sim$45 than the longest
risetime seen from this source to date. The peak flux calculated over a
64-ms time interval (or over the risetime interval) is the lowest of all
four bursts. The total burst fluence is within the range of fluences
observed for the other bursts from this source.
%%% bursts 1 2 3 4
%%% risetime 21 ms 6 ms 18 ms 43 ms
%%% t90 51 s 7 s > 700 s 6000 s
%%% fluence 20 5.3 >330 77 e-10erg/cm2 (TIMESCALE?x?)
%%% peak flux 31/54  26/114   59/105   42  e-10erg/s/cm2 (TIMESCALE? 1st num 64 ms % second risetime)

\begin{figure*}
\begin{center}
\includegraphics[scale=1.1]{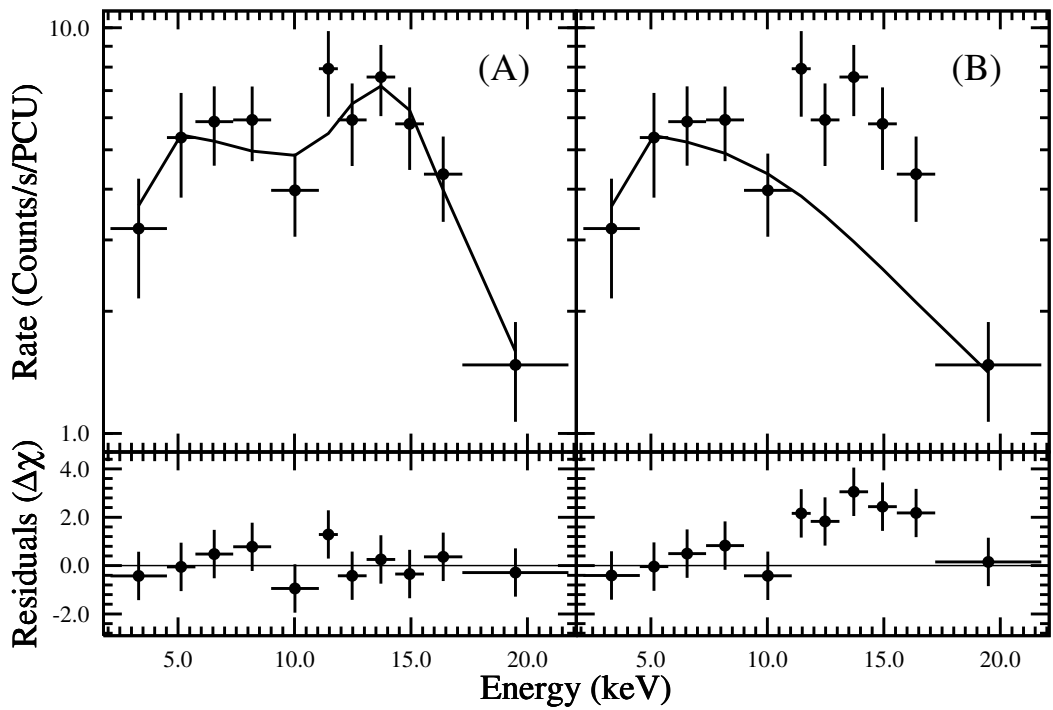}
\caption
%%{Spectral fits of the first 3 seconds of the new burst. {\emph{Top:}}
%%Photoelectrically absorbed power-law model. {\emph{Bottom:}}
%%Photoelectrically absorbed power-law plus Gaussian model.
{Spectral fits of the first 3 seconds of the new burst.  ({\emph{a}})
Photoelectrically absorbed power-law plus Gaussian model.
({\emph{b}}) Photoelectrically absorbed power-law model.
\label{plot-spectrum}}
\end{center}
\end{figure*}

%% spectral parameters %%
A burst spectrum was extracted using all the counts above 2~keV within the
$T_{90}$ interval. The spectrum was then grouped so as to have at
least 20 counts per bin after background subtraction. The bin above $\sim$40~keV was
ignored because it had insufficient counts even after grouping. A response
matrix was created using the \texttt{FTOOL} \texttt{pcarsp}. The burst
spectrum, background spectrum, and response matrix were then read into
\texttt{XSPEC}. 
The spectrum was fit to a photoelectrically absorbed blackbody and to
a photoelectrically absorbed power-law.  In both cases, because of
\emph{RXTE}'s lack of response below 2~keV, we held the column density
fixed to the value found by \cite{tgd+08}
(N$_H$~=~0.97~$\times$~10$^{22}$~cm$^{-2}$).
%%%% using \textit{Chandra X-ray Observatory}.  
The power-law model was a poor fit. 
The blackbody
was a better fit but not exceptional, with a 25$\%$
probability that the deviations from the model are due
to random noise only, given the number of
degrees of freedom (see Table~\ref{table:burst}).
%
%The blackbody
%was a better fit but not exceptional, having a 25$\%$ chance of
%exceeding the observed $\chi^2$ value given the number of degrees of
%freedom (see Table~\ref{table:burst}). 
%
Two component-models such as
two blackbodies or a blackbody plus power-law did not improve the
fits.
Two other bursts from this source exhibited spectral features at
$\sim$14~keV. 
To determine whether the burst exhibited any spectral features that
might have been smeared out by
extracting a spectrum over the burst's long $T_{90}$ interval, we
repeated the above procedure for the first few seconds of the burst.
%
% [[[[[[[[[ CLEARNESS ]]]]]]]]]
There was indeed excess at $\sim14$~keV when fitting the spectrum of
the burst with a simple continuum model (see
Fig.~\ref{plot-spectrum}). To establish the veracity of the feature we
performed the following Monte Carlo simulations.  
We generated 1000 simulated spectra having the same count rate and
exposure as our data and a photoelectrically absorbed blackbody shape.
The simulated spectrum was created using the $kT$ value found from the
best fit line+gaussian model. We then added Poisson noise to our
simulated spectra, fit them with a simple photoelectrically absorbed
blackbody model, and calculated a $\chi^2$ value. Next, we added a
spectral line to our fitting model and refit the data. To prevent the
fit from falling into a local minimum, when doing the fitting, we
stepped the central energy of the line from 2~keV to 40~keV in steps
of 0.1~keV, but allowed the width and normalization of the line to
vary. We found 17 cases out of 1000 for which the addition of a line
induced a change in $\chi^2$ that was greater than or equal to that found
from adding a line to the model used to fit the real data.
Thus, we place a significance of 99.983\% on the line which is
equivalent to a 2.1-$\sigma$ detection. This is not a highly
significant detection, nor is it as significant as the other lines
seen from this source \citep{gkw02, gkw06}, however, it is highly suggestive that given
there have been only four bursts seen from this source, two of which have
shown similar features at comparable energies.

\subsection{Short-Term Pulsed Flux Variability and Burst Phases}

\begin{figure*}
\begin{center}
\includegraphics[scale=0.75]{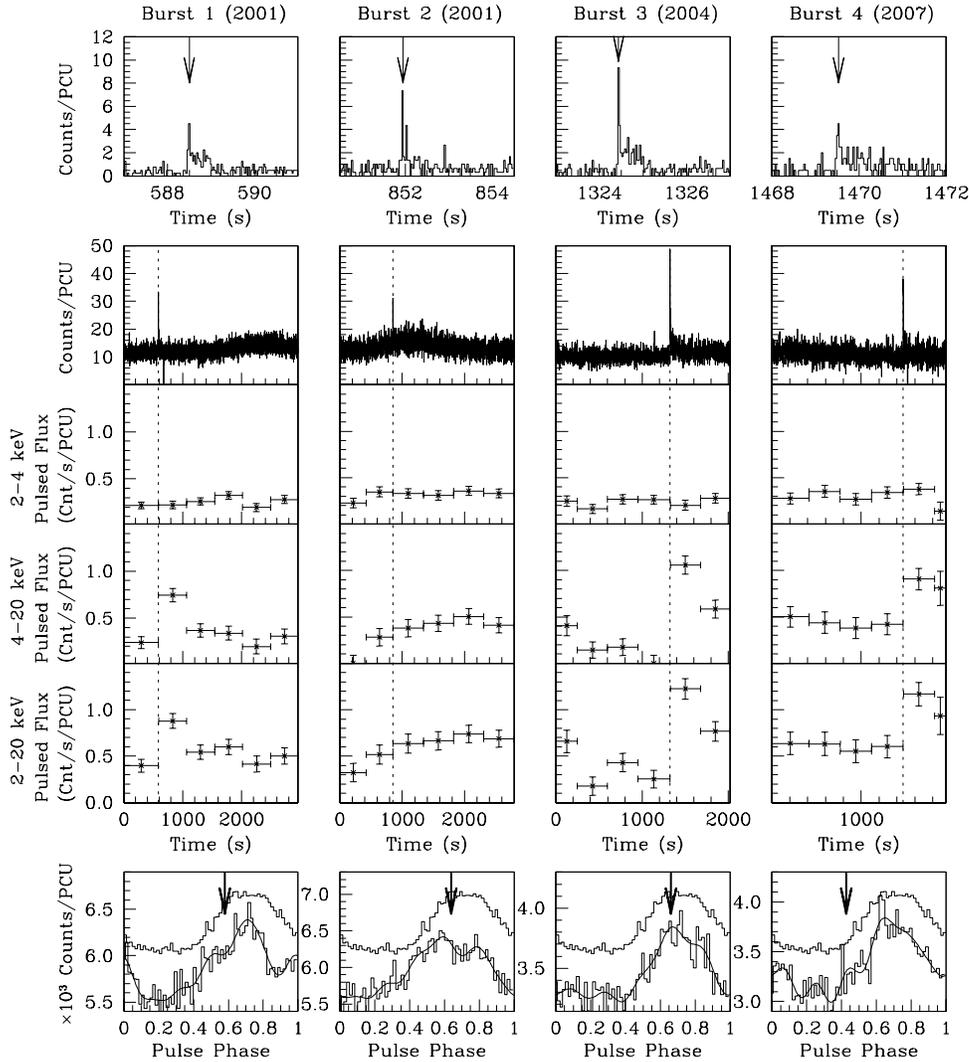}
\caption
{Short-term pulsed flux variability near the bursts, and burst phases.
Each column corresponds to an observation
in which a burst was detected. In each column: {\emph{Top:}} a 4-s long time
series with 31.25~ms time resolution showing the burst. The peak of the burst
is indicated with an arrow. {\emph{Middle:}} time series of the entire
observation with 1-s time resolution, followed by the RMS pulsed flux in
three different bands. We excluded the 4~s surrounding the burst
for this pulsed flux analysis. {\emph{Bottom:}} A fold of the entire
observation shown below the scaled long-term average profile. The phase
at which the burst occured is marked with an arrow. This phase corresponds
to the time bin indicated with an arrow in the top plot.
\label{plot-bursts}}
\end{center}
\end{figure*}

%% pulsed flux %%
We show the 2$-$20~keV lightcurves of the 4 seconds surrounding each of the
4 bursts discovered from 1E~1048.1$-$5937 in the top panels of
Figure~\ref{plot-bursts}. Each column corresponds to a burst. For each of
the observations containing bursts, we made three barycentric time series in
count rate per PCU, for the 2$-$4, 4$-$20, and 2$-$20~keV bands. The time
resolution was 31.25 ms. We removed the 4~s centered on each burst from each
time series. Then, we broke each time series into six segments. For each
segment, we calculated the RMS pulsed flux. The results are presented in the
middle panels of Figure~\ref{plot-bursts}. A similar analysis was performed
for burst~3 by \cite{gkw04}. Note the significant increase in the 4$-$20~keV
pulsed flux in the observations containing the first, third, and last bursts
following the onset of the bursts, while the pulsed flux remained constant
in the 2$-$4~keV band in those same events.

%% phases %%
To determine the phase of each burst, we folded each observation at the
best-fit frequency and found when, relative to the folded profile, the burst
peak occured. We then aligned each of the folded profiles with a high
signal-to-noise long-term average profile. The pulse phases are shown in the
bottom panels of Figure~\ref{plot-bursts}. In each panel, the histogram at
the bottom is a fold of the entire observation. The smooth curve is obtained
from the best-fit 5 harmonics. The histogram at the top is the long-term
average. The first, second, and fourth folded profiles have a different
shape from the long-term average; they occured during flares. Note how the
first three bursts occur near the peak of the profile (burst phases
0.58$\pm$0.02, 0.64$\pm$0.02, and 0.66$\pm$0.02 relative to the template
shown in the Figure), but the last burst is further from the peak (burst
phase 0.43$\pm$0.02 relative to the template shown in the Figure).

\section{Discussion}
\label{sec:discussion}

\subsection{Pulsed Flux Variations}
\label{sec:discussfluxes}

The goal of continued systematic {\emph{RXTE}} monitoring of AXPs is to
flesh out the phenomenological phase space of these intriguing objects. In
this regard, 1E~1048.1$-$5937 has not disappointed us. It has shown a
surprisingly diverse range of behaviors in practically every observational
property. This includes its rotational evolution, in which we have seen
several different timing anomalies -- with two likely spin-up glitches -- in
addition to remarkable timing ``noise'', for lack of a better term.  Its
radiative evolution has been equally eventful, with 3 large, long-lived
pulsed flux increases and multiple bursting episodes, as well as spectral
changes and pulse profile changes.  Understanding the physical origin of all
this behavior is clearly very challenging; likely the best physical insights
will come from considering multiple studies such as ours, for many different
objects. Nevertheless here we consider what these phenomena may be telling
us about the physics of magnetars.

%%% [[LINEBREAK APJ]]
Several AXPs have exhibited pulsed flux variations on long timescales. \linebreak
RXS~J170849.0$-$400910 exhibited low-level pulsed flux variations on
timescales of weeks to months \citep{dkg08}. 4U~0142+61 exhibited a pulsed
flux increase by 29$\pm$8\% over a period of 2.6 years \citep{dkg07UPDATED}.
1E~2259+586 exhibited an abrupt increase in the pulsed (and persistent) flux
which decayed on timescales of months to years \citep{wkt+04}. This abrupt
increase occured in conjunction with bursts, and the decay is thought to be
due either to thermal radiation from the stellar surface after the
deposition of heat from bursts (eg. \citeauthor{let02} \citeyear{let02}), or
the result of the slow decay of a magnetospheric ``twist'' \citep{tlk02}.
This outburst was accompanied by a glitch. XTE~1810$-$197 and candidate AXP
AX~J1845$-$0258 also exhibited an increase in the pulsed flux although the
risetime is unclear \citep{ims+04,tkk+98,tkgg06}.

The long-term pulsed flux behavior of AXP 1E~1048.1$-$5937 is different from
that of any other AXP: in the first two flares exhibited by
1E~1048.1$-$5937, the pulsed flux rose on week-long timescales and
subsequently decayed back on time scales of months to years \citep{gk04}. It
is unclear whether the third flare had a resolved rise (see
Section~\ref{sec:flux}). Although small bursts sometimes occured during
these events (see Section~\ref{sec:bursts}), the afterglow of these small
bursts cannot explain the overall flux enhancement, and in the absence of
evidence for large bursts prior to the flare, we can attribute the flares to
twists implanted in the external magnetosphere from stresses on the crust
imposed by the internal magnetic field.

Based on the idea that a plasma corona is contained within the closed
magnetosphere, \cite{BT07} offer a prediction for the behavior of the
luminosity of the source after a magnetospheric twist occurs. Assuming that
a large flux enhancement is caused by a twist, that the emission from the
heated crust is small compared to the magnetospheric emission of twisted
magnetic flux tubes, and assuming no additional twists introduced after the
original twist, Equation~17 of \cite{BT07} predicts that the luminosity will decay
linearly, and is proportional to $-{{\phi}^{2}} \times (t-t_0)$, where ${{\phi}}$
is the voltage between two footpoints of a magnetic field line and
$(t-t_0)$ is the time since the start of the decay of the luminosity. ${\phi}$ is induced
by the current that accompanied the gradual untwisting of the magnetic
field. Its minimum value is that needed for the creation of electron
positron pairs. It is proportional to the local magnetic field.

\begin{figure*}
\begin{center}
\includegraphics[scale=0.75]{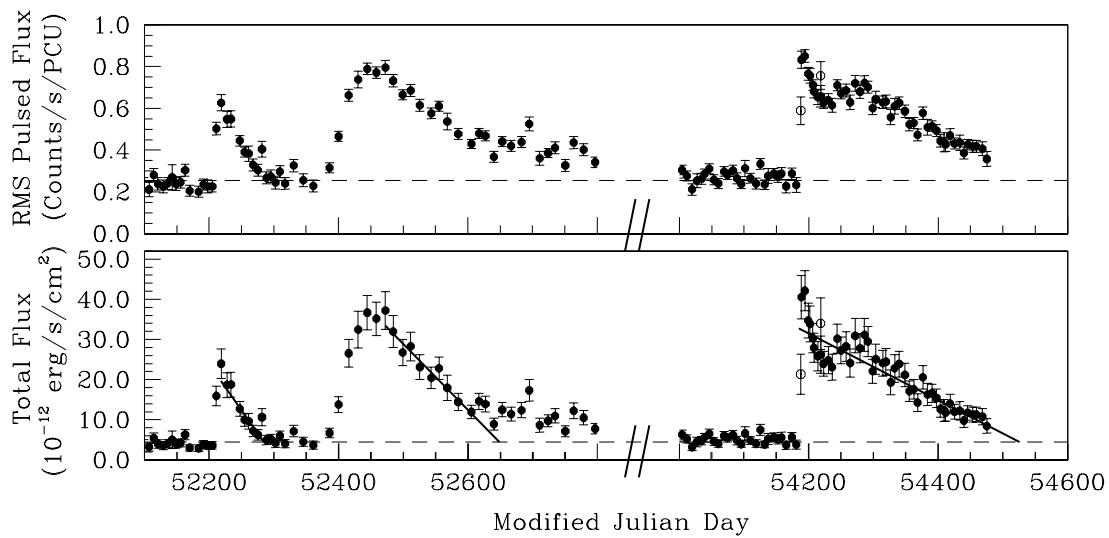}
\caption
{{\emph{Top:}} Pulsed flux in the 2$-$10~keV near the three flares.
{\emph{Bottom:}} Simulated total 2$-$10~keV unabsorbed flux, estimated from
the {\emph{RXTE}} pulsed flux and from the power-law correlation between the
pulsed fraction and the total flux described by \cite{tgd+08}. The solid
lines in the bottom plot are linear decays fit to the first few months of
data after each of the flares.
\label{plot-linearfit}}
\end{center}
\end{figure*}

%%% [[LINEBREAK ASTROPH]].
Using Equation~\ref{eq:eq4} and our pulsed flux time series, we produced a
total flux time series for 1E~1048.1$-$5937 (Figure~\ref{plot-linearfit}).
We then fit a linear decay to the first few months of data after each of the
three flares. Including data beyond that would have made the fits worse,
indicating that we can attribute at most the first part of the decay to a
linear twist relaxation. The second part could perhaps be attributed to crust
afterglow following some internal heat deposition (see
Section~\ref{sec:discussglitches}). The slopes of the linear fits had the
values $-$0.23(3)$\times 10^{-12}$~erg/s/cm$^2$/day, $-$0.164(13)$\times
10^{-12}$~erg/s/cm$^2$/day, \linebreak and $-$0.083(5)$\times
10^{-12}$~erg/s/cm$^2$/day, with reduced ${\chi}^{2}$ values of 0.72, 0.38,
and 0.91, respectively. The fits were good, as predicted by \cite{BT07}, but
the three slopes significantly differed from each other, suggesting, in the
context of this model, that different flux tubes (with different values of
local magnetic fields) were twisted in each event.

\subsection{Timing Behavior}

In addition to flux variability, regularly monitored AXPs also exhibit
different kinds of timing variability. In RXS~J170849.0$-$400910, the
frequency derivative fluctuates by $\sim$8\% about its long-term
average $\sim$1.58~$\times 10^{-13}$~s$^{-2}$ on a timescale of months
\citep{PREPGLRIM}, except at the second detected glitch which had an exponential
recovery \citep{kg03}. The frequency derivative of 1E~1841$-$045 varies by
$\sim$10\% on a timescale of many years, except at the first detected glitch
where $\dot{\nu}$ suddenly dropped by $\sim$10\% \citep{dkg08}. It slowly
dropped further before slowly recovering. The frequency derivative of
4U~0142+61 also fluctuates by $\sim$3\% around its long-term average on a
timescale of months to years, except at the onset of the 2007 active phase
where it suddenly dropped \citep{gdk09submitted}. It then slowly
recovered. The frequency derivative of 1E~2259+586 also fluctuates about its
long-term average, except at the first detected glitch, which had an
exponential recovery \citep{wkt+04}.

The episode of extreme variations in $\dot{\nu}$ of 1E~1048.1$-$5937 is not
seen in any other AXP. In 2002 and 2003, $\dot{\nu}$ varied by
$\sim$5.6$\times10^{-13}$~s$^{-2}$ every two weeks (time between consecutive sets
of three observations), oscillating between $-$26$\times 10^{-13}$~s$^{-2}$ and
$-$6$\times 10^{-13}$~s$^{-2}$ 4 times in the span of 450 days.

\begin{figure*}
\begin{center}
\includegraphics[scale=0.75]{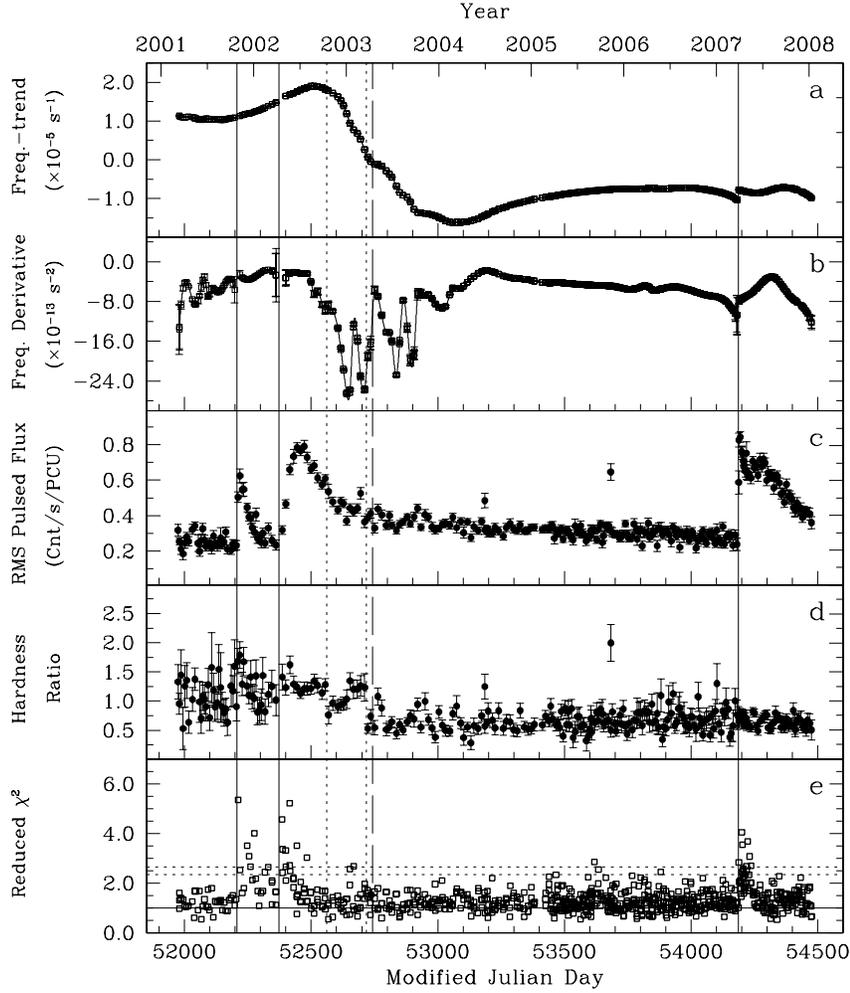}
\caption {Evolution of the different properties of 1E~1048.1$-$5937.
({\emph{a}}) Frequency as a function of time, with the
long-term average spin-down subtracted. ({\emph{b}}) The frequency
derivative as a function of time. ({\emph{c}}) The 2$-$10~keV RMS pulsed
flux as a function of time. ({\emph{d}}) The hardness ratio as a function
of time, computed from the pulsed flux in the energy range
(4$-$10~keV)/(2$-$4~keV). ({\emph{e}}) Reduced ${\chi}^{2}$ statistics as
a function of time, calculated after subtracting the scaled and aligned
profiles of the individual observations from a high signal-to-noise
template. {\emph{All panels:}} The three solid lines mark the onset of the
three flares. The line with the long dashes marks the location where the
ephemerides used to obtain the splines overlapped for a short period of time
only. The two dotted lines mark epochs where the hardness ratio dropped, and
maintained a lower value for the following weeks relative to the values
preceding the drop.
\label{plot-compare}}
\end{center}
\end{figure*}

When trying to understand the origin of these variations, it is useful to
look for correlation between the timing properties and the flux of the
pulsar. Figure~\ref{plot-compare} is a plot of the timing and
radiative behaviors of 1E~1048.1$-$5937.
%%% with solid lines indicating the
%%% onsets of the flares, dotted lines indicating drops in the hardness ratio,
%%% and a dashed line indicating a particularly noisy epoch. 

Earlier we suggested that the flux variations in 1E~1048.1$-$5937 may be due to twists implanted in
the external magnetosphere from stresses on the crust imposed by the
internal magnetic field. In the magnetar model, the twisting drives currents
into the magnetosphere. The persistent non-thermal emission of AXPs is
explained in this model as being generated by these currents through
magnetospheric Comptonization \citep{tlk02}. Changes in X-ray
luminosity, spectral hardness, pulse profile, and torque changes have a
common origin in this model.

\cite{gk04} looked for correlations between the $\dot{\nu}$ and the pulsed
flux near the first two flares and reported only a marginal correlation.
They suggested that the lack of correlation was because the torque is most
sensitive to the current flowing on a relatively narrow bundle of field
lines anchored close to the magnetic poles \citep{tlk02}. Therefore, whether
an X-ray luminosity change will be accompanied by a $\dot{\nu}$ change
depends on where in relation to the magnetic pole the source of enhanced
X-rays is.

Earlier we suggested that the different decay slopes of the three different
flares might indicate that different flux tubes, with different values of
local magnetic field, were twisted in each event \citep{BT07}. Therefore,
even if there was no correlation between $\dot{\nu}$ and the pulsed flux in
the case of the early flares, if the flux tubes involved in the third flare
were closer to the poles, one might expect a correlation to occur in that
flare. From Figure~\ref{plot-compare}, this does not appear to be the case.
Note that correlations between the luminosity and torque are also expected
in accreting scenarios, and are not observed here.

An interesting observation is that episodes of rapid $\dot{\nu}$ variations
appear to follow the second and third flares (see Section~\ref{sec:flux}).
Bi-monthly variations in $\dot{\nu}$ changed from
0.08$\times10^{-13}$~s$^{-2}$ near the second flare to
2$\times10^{-13}$~s$^{-2}$ 41 days after the same flare, to
5.6$\times10^{-13}$~s$^{-2}$ 141 days after the flare. Similarly, weekly
variations in $\dot{\nu}$ changed from 0.33$\times10^{-13}$~s$^{-2}$ near
the third flare to 0.6$\times10^{-13}$~s$^{-2}$ 350 days after the flare, to
1.4$\times10^{-13}$~s$^{-2}$ 395 days after the flare. This might only be a
coincidence, however \cite{BT07} predict that the impact of a twist in the
magnetosphere on the spin-down may appear with a delay of $\sim$2~years. This
is because the timescale of the twist spreading to the light cylinder is
large due to the resistivity to the currents of the corona contained within
the closed magnetosphere.

\subsection{Glitches}
\label{sec:discussglitches}

In many glitch models, the superfluid in the crust is spinning faster than
the crust, but on average over long times, they have the same $\dot{\nu}$.
The superfluid cannot spin down because its angular momentum vortices are
pinned to crustal nuclei and hence cannot move outward (see, for example,
\citeauthor{acp89} \citeyear{acp89}). For various reasons, for example,
torques on the crust, internal starquakes and thermal agitations, unpinning
of the vortex lines may happen in some locations. The vortices could then
move outward, and the superfluid angular frequency can decrease and approach
that of the crust. At that moment, angular momentum is transferred from the
superfluid to the crust, and a glitch occurs.  For example, \citet{le96}
argued that starquakes due to magnetic stresses at the core/crust boundary
in normal rotation-powered pulsars could deposit energy that then results in
sudden spin-ups; such events seem even likelier to occur in magnetars,
consistent with their ubiquitous glitching
\citep{klc00,kg03,dis+03,dkg08,igz+07}.

The glitch coincident with the third pulsed flux flare of
1E~1048.1$-$5937 is the largest yet seen in the five regularly monitored AXPs,
and has one of the largest fractional frequency increases in any pulsar,
including rotation-powered sources. This event was not accompanied by a significant
change in $\dot{\nu}$. In fact, in the months preceding the glitch,
$\dot{\nu}$ became more and more negative, until three weeks prior to the
glitch, when it started decreasing in magnitude, reaching a value not far
from the one it adopted after the glitch. It is unclear if this behavior is
somehow related to glitch; perhaps it caused the unpinning of the vortices.
Unfortunately, the relation between what is usually considered to be timing
noise and the behavior of the superfluid inside is not well understood.

%%% What is clear, however, is that this glitch was associated with a radiative
%%% event: the third pulsed (and persistent) flux flare. Above, we suggested
%%% this flare may be due to a twist in the magnetosphere. The twist originates in a tangle
%%% of field lines below the surface of the star. Because of the internal
%%% magnetic stresses inside, a piece of crust above the tangle is twisted,
%%% twisting the footpoints of the external magnetic field. Eventually this
%%% twist propagates outward. It is possible that some vortices that are
%%% pinned to the crust get dislodged when the crust is being twisted, causing a
%%% glitch.
What is clear, however, is that this glitch was associated with a radiative
event: the third pulsed (and persistent) flux flare. Above, we suggested
this flare may be due to a twist in the magnetosphere. The twist originates
in a tangle of field lines below the surface of the star. Because of the
internal magnetic stresses inside, a piece of crust above the tangle is
twisted, twisting the footpoints of the external magnetic field. Eventually
this twist propagates outward. It is possible that some vortices that are
pinned to the crust get mechanically dislodged when the crust is being
twisted, causing a glitch, or that energy deposition during this event
raises the temperature such that pinning is affected, as in the \citet{le96}
picture.

In fact, we note that every observed AXP flare or outburst thus far has been
accompanied by a timing event. 
%%%%%%%%
%In the case of 1E~2259+586,
%CXOU~J164710.2$-$455216, and the third flare of 1E~1048.1$-$5937, the event
%was a glitch (\citeauthor{wkt+04} \citeyear{wkt+04}, \citeauthor{icd+07}
%\citeyear{icd+07}, and Section~\ref{sec:timf3}). 
%%%%%%%%
In the case of 1E~2259+586 and the third flare of 1E~1048.1$-$5937, the
event
was a glitch (\citeauthor{wkt+04} \citeyear{wkt+04}, and Section~\ref{sec:timf3}). 
%%%%%%%%
For 4U~0142+61 and for the
second flare of 1E~1048.1$-$5937, the event was a sudden change in
$\dot{\nu}$ possibly accompanied by a glitch (\citealt{gdk09submitted} and
Section~\ref{sec:timf2}). For the first flare of 1E~1048.1$-$5937 the event
was a timing anomaly of uncertain nature (Section~\ref{sec:timf1}). It is
possible that all these timing events were caused by some unpinning of
superfluid vortices, which in turn was caused by crustal movement due to a
twist propagating outward.

Note however that the converse is not true: many AXP glitches appear to be
radiatively silent, such as the second glitch observed from 1E~2259+586, and
all 4 glitches of AXP 1E~1841$-$045 (\citealt{dkg08}, 
\citealt{PREPGLRIM}).  There is no evidence of pulsed flux changes
associated with the glitches of RXS~J170849.0$-$400910, however there are
claims of an association between variations in the total flux of the source
and the glitch epochs (see, for example, \citeauthor{cri+07}
\citeyear{cri+07} and \citeauthor{igz+07} \citeyear{igz+07}). 
%%% % This is puzzling because when angular momentum is transferred, slowing down the
%%% % superfluid and spinning up the crust, there is a loss of total rotational
%%% % kinetic energy ** VICKY: YOU SAID TO ESTIMATE THIS LOSS. I DID. BUT I HAVE
%%% % QUESTIONS ABOUT THE CALCULATION. This energy is released somewhere,
%%% % presumably internally. 
%%% % Why do we not we see it in the case of the ``silent'' glitches?
%%% Why do we not see any released energy in the case of the `silent'' glitches?
%%% % Perhaps the superfluid is efficient at redistributing the thermal
%%% % energy inside the star. Or, 
%%% Perhaps it is released deep in the star and some, or all, of it does not
%%% reach the surface. Perhaps when it reaches the surface, the risetime of the
%%% associated increase in the flux is related to how deep below the surface
%%% this energy was released (see, for example, \citeauthor{am04}
%%% \citeyear{am04} and BROWNANDCUMMING2008.
Why do we not see any released energy in the case of the ``silent''
glitches? Perhaps it is released deep in the star and some, or all, of it
goes into the core. Perhaps when some energy reaches the surface, the delay
between the energy release and the start of the associated increase in flux
as well as the actual risetime of that increase are related to how deep
below the surface this energy was released (see, for example,
\citeauthor{am04} \citeyear{am04}, \citeauthor{bc08} \citeyear{bc08}). Since
the risetime of the flux events associated with the ``loud'' glitches is
never longer than a few months, one could speculate that any energy release
that would have caused a larger risetime goes directly into the core.

Indeed, perhaps the radiative events accompanying some glitches are not due
to a twist of the footpoints of the external field following crustal
cracking. Perhaps the sudden unresolved increases in the flux, like that
seen in the 2002 event from 1E~2259+586, are due to a twist propagating from
the inside by breaking the crust (possibly combined with a thermal energy
release due to the glitch), while the slow resolved increases in the flux,
like those seen in the first two flares of 1E~1048.1$-$5937, are due to a
local thermal energy release following a glitch.

\subsection{Pulse Profile Changes}

In Section~\ref{sec:profiles}, we showed that the largest pulse profile
changes happen near the flares. In 1E~1048.1$-$5937, these changes always
involve an increase in the harmonic content of the profiles. This suggests
that these changes are not due to a surface disturbance (hot spot), since
the effects of this on the profile would probably be smeared due to general
relativistic light bending \citep{dpn01}. Instead, they may be due to a
local event in the magnetosphere.

\cite{BT07} argued that it might take several years for the twist to
propagate from the surface of the star to the light cylinder. Once the twist
reaches the light cylinder, the torque affects the star almost immediately,
and variations in $\dot{\nu}$ are observed.
%%% This torque immediately affects the star (because the alfven crossing time
%%% usual magnetospheric disturbances not involving twists is very short) -zibarth
In this picture, early on, the twist is in the lower magnetosphere, and much
later it is in the upper magnetosphere. A twist in the lower magnetosphere
where the fields are very strong affects the properties of the local plasma,
modifying its emission as well as the emission that is scattered
from the surface below. In this case one might expect noticeable pulse
profile changes. A twist that has reached the light cylinder where the
fields are weaker affects the properties of the local plasma less, and
affects the emission from the lower magnetosphere less. In this case one
might expect the pulse profile changes to be much smaller. Thus, that we are
seeing the pulse profile changes only at the beginning of the flares is
consistent with the picture in \cite{BT07}.

If significant pulse profile changes {\emph{had}} been observed at the time
of the large $\dot{\nu}$ variations, this could have meant that the pulsed flux
flares and the subsequent large $\dot{\nu}$ variations are independent. In
this case, the $\dot{\nu}$ changes would be due either to a low
magnetospheric twist (accompanied by pulse profile changes) which propagates
quickly to the light cylinder, causing the torque changes; but why the
onset of this twist would not be accompanied by a visible energy release is
unclear. Alternatively, the $\dot{\nu}$ changes may not be due to torques at
the light cylinder, but to internal events which cause the crust to crack
and the lower magnetosphere to twist itself.
%%% , causing the pulse profile to change. 
In this case too it would be unclear why no bursting activity or energy
release was seen,
and the unobserved pulse profile changes would be puzzling.

\subsection{Long-Term Spectral Changes}

In Section~\ref{sec:flux} we show that
the hardness ratio of 1E~1048.1$-$5937, obtained from the pulsed flux,
dropped significantly 9 months after the peak of the second flare, while the
pulsed flux was still decaying, and while the large $\dot{\nu}$ variations
were ongoing. Unfortunately, there are no imaging observations of the source
around that time, and we cannot verify if the drop in the pulsed hardness
ratio was accompanied by a drop in the hardness ratio of the persistent
emission. Assuming that it was, this softening may be related to the
magnetospheric twist that caused the flares.  Indeed in the \cite{tlk02}
model, spectral hardness is correlated with the luminosity. The hardness
ratio is expected to gradually drop when the flux decays.
However, the decrease seen here was sudden, not gradual. Also, this does not
explain why the hardness ratio was not lower prior to the onset of the
flares. Alternatively, this correlation of the softening with the flux decay
can also be a consequence of changes in the effective temperatures of the
outer layers of the star \citep{og07check}, and the twisted magnetosphere model
need not be invoked here.

\subsection{Bursts}
\label{sec:discussbursts}

1E~1048.1$-$5937 is more active than the other AXPs we monitor in multiple
regards. With {\emph{RXTE}} monitoring over the past $\sim$10~years, it is
the only one which exhibited three large flux flares, it is the only
AXP which exhibited extreme variations in $\dot{\nu}$, and the glitch
observed in conjunction with the third flare is the largest observed among
these AXPs. In addition, 1E~1048.1$-$5937 has shown four bursts, at
different epochs. Other AXPs, such as RXS~J170849.0$-$400910 and
1E~1841$-$045, have shown none, even though the combined {\emph{RXTE}} on-source time of these
two sources is the same as that of 1E~1048.1$-$5937.

%%% I haven;t compared to AXPs 2259 and 0142.
%%% 1048: at least 896*2ks = 1792
%%% 1708: at least 360*2ks = 720
%%% 1841: at least 192*5ks = 960
%%% 2259: at least 389*5ks = 1945

%%% To date, four bursts have been detected from this source, with the
%%% possibility of a fifth burst having occured just before the the observation
%%% with the unusually high 4$-$10~keV pulsed flux at the end of 2005 (see
%%% panel~3 of Figure~\ref{plot-flux}). 
Since three of the observed bursts were followed by an enhancement in the
4$-$10~keV pulsed flux, it is possible that the high pulsed flux of the
observation taken on 2005 November~08 (see Section~\ref{sec:flux}) was due
to a burst that occured just before the observation. We detected no evidence
for a change in pulsed flux during this observation. Since it was the first
in the weekly set of three closely spaced observations, the decay timescale
of the putative burst could be such that the decay is not noticeable within
the observation, given the size of our pulsed flux uncertainties. The
observation that followed this one occured 18 hours later and its pulsed
flux was consistent with the long-term average. We
verified that there were no {\emph{SWIFT}} triggers from the location of
1E~1048.1$-$5937 in the week preceding the anomalous observation.

The four bursts observed from 1E~1048.1$-$5937 were associated with
different flaring events. Bursts~1 and~2 occured near the peak of the first
flare. Burst~3 occured two years after the peak of the second flare, while
the pulsed flux was still decaying. Burst~4 occured a month after the peak
of the third flare. Bursts~1, 2, and~3 occured near pulse maximum. All four
bursts had millisecond risetimes. Bursts~1, 3, and~4, had long decay tails
(51~s, $>$~700~s, and 128~s), with a pulsed flux enhancement in the tails.
The falltime for burst~2 was 2~s. An apparent feature near 13~keV has been
observed in the spectra of the first, and third bursts \citep{gkw02,gkw04},
and an apparent feature near 15~keV has been observed in the spectrum of the
most recent burst. Note that apparent features near 13~keV have also been
observed in the tail of one of the bursts in AXP~XTE~J1810$-$197
\citep{wkg+05}, and in the spectrum of the largest burst detected in
AXP~4U~0142+61 \citep{gdk09submitted}. So far the presence of these
features is not well understood.

\cite{wkg+05} suggest that there are two types of magnetar bursts.
Type~A bursts are short, symmetric, and occur uniformly in pulse phase.
Type~B bursts have long tails, thermal spectra, and occur preferentially at
pulse maximum. They also noted that Type~B bursts occur preferentially in
AXPs (although AXP 1E~2259+586 emitted both kinds of bursts during its 2002
outburst), and Type~A bursts occur primarily in SGRs.

\cite{wkg+05} argue that type~A bursts are due to reconnections in the upper
magnetosphere, and that type~B bursts are due to crustal fracture followed
by a rearrangement of the magnetic field lines outside the surface. They
explain that a magnetospheric origin would lend itself to more isotropic
emission having no preference for a particular pulse phase, while the crust
fracture model would naturally produce a phase dependence of the burst
emission for a localized active region on the crust. The tendency of the
bursts to occur near pulse maximum is consistent with the strain in the
crust causing the cracking being highest in the regions where the field is
the strongest: at the polar caps. Furthermore, \cite{tlk02} have argued that
the SGRs, with their strong non-thermal spectral components, undergo more
reconnection events. Therefore, if the type~A bursts are really due to
magnetospheric events, then it makes sense that they occur more in SGRs.

%%% [[LINEBREAK ASTROPH]].
None of the bursts observed from 1E~1048.1$-$5937 is of the symmetric type.
All four bursts had a long tail with pulsed flux enhancement, except in
burst~2 where the tail was very short, but still $\sim$~300 times longer
than the risetime. Therefore all four are probably bursts of Type~B.
However, burst~4 did not occur near pulse maximum. This does not necessarily
mean that the above interpretation of Types~A and~B bursts is wrong; rather,
for this burst, perhaps the crustal cracking did not occur near the polar
cap, or near the hot spot that usually yields the pulse.

Note that a similar situation occured for AXP~4U~0142+61. Six bursts were
detected from this AXP during the 2006 active phase
\citep{gdk09submitted}. None of them was a short and symmetric Type~A burst: they all had
tails, although in two cases the tails were shorter than 10~s. Burst~1
occured at pulse maximum. Bursts~2 to~5 all occured within a single
observation, and two of them did not occur near pulse maximum. Presumably
some global event had caused the crust to crack at many places. 
%%% Burst~6 is one of the most energetic AXP bursts to date.
Not only did burst~6 not occur at pulse maximum, but it occured where a
temporary new peak in the profile appeared. Here, just like for burst~4 of
1E~1048.1$-$5937, a large crack could have appeared away from the usual
location of the emission.

\section{Summary}

We have presented a long-term study of the timing properties, the
pulsed flux, and the pulse profile of AXP 1E~1048.1$-$5937 as measured
by {\emph{RXTE}} from 1996 to 2008. We showed that the onset of the
2001 pulsed flux flare was accompanied by a timing anomaly and by
significant pulse profile changes. 
%
%%% Assuming no systematic drift in the phase of the pulse profile,
%
The timing anomaly was consistent
with a gradual slow down lasting 2$-$3 weeks followed by a recovery.
We showed that the onset of the 2002 pulsed flux flare was accompanied
by a likely glitch of size
$\Delta\nu/\nu$~=~2.91(9)~$\times$~10$^{-6}$, by a large change in
$\dot{\nu}$~=~$-$4.10(14)~$\times$~10$^{-14}$, and by significant
pulse profile changes. 
%
%%% We assumed no systematic drift in the phase of the pulse profile when
%%% we derived the glitch parameters.
%
We use the term ``likely'' because, while the
trend in the timing residuals indicates that a glitch occured, due to
the finite resolution of the data, which is particularly problematic
given the extreme timing noise of this source, a rapid
non-instantaneous variation cannot be ruled out. 
%
%The timing results reported above for the first two flares assume no
%systematic drift in the phase of the pulse profile.
%
%[[[For these two flares,
%contamination of the timing results from a systematic drift in the
%phase of the pulse profile cannot be completely excluded.]]] 
%
Both of
these flares had few-weeks-long risetime. 
Several months after the
peak of the second flare, and while the pulsed flux was still
decaying, the source underwent extreme $\dot{\nu}$ variations lasting
$\sim$450 days. Then the source entered a period of relative timing
quiescence in which no radiative changes were observed except for
occasional low-level pulse profile changes. The source reactivated in
2007 and a third pulsed flux flare was observed. The risetime of that
flare was $<$~7.3~days. It is unclear whether the rise was resolved.
Contemporaneous imaging observations showed that the persistent flux
rose also. The onset of this flare was accompanied by a very large
spin-up glitch ($\Delta\nu/\nu$~=~1.63(2)~$\times$~10$^{-5}$) and by
many significant but short-lived pulse profile changes. In total, four
short non-symmetric bursts have been observed in this source to date.

The three pulsed flux flares can be attributed to twists implanted in the
external magnetosphere from stresses on the crust imposed by the internal
magnetic field. \cite{BT07} postulated the presence of a plasma corona
within the closed magnetosphere and predicted a linear decay in the flux
following the initial rise due to the twist. The first part of the decay of
the observed flares can be well fit with a linear trend, but not the entire
decay. Alternatively, the flares can be attributed to an internal heat
release associated with the contemporaneous timing events, although the
pulse profile changes seen contemporaneously with the flares likely have a
magnetospheric origin. All three flares were accompanied by either a timing
anomaly or a glitch. This can be due to a disturbance in the superfluid
vortex lines caused by the crustal disturbance at the time the twist was
implanted. The extreme timing noise observed several months after the peak
of the second flare may be attributed, in the \cite{BT07} picture, to the
twist associated with the flare finally having reached the light cylinder,
although it is hard to understand the magnitude and the timescale of the
variability in this picture. Finally, all four bursts observed in this
source can be attributed to the crustal cracking that occured when the twist
propagated from the inside of the star to the lower magnetosphere.

A coherent physical picture explaining the variety of behaviors observed in
this fascinating source, as well as in other AXPs, has yet to emerge,
however we hope through continued detailed studies such as the one presented
here, one will be forthcoming soon. Thus far, the framework of the magnetar
model appears most promising to us.

\acknowledgments

We thank A. Cumming and D. Eichler for useful discussions. Support was
provided to VMK by NSERC Discovery Grant Rgpin 228738-08, an FQRNT Centre
Grant, CIFAR, the Canada Research Chairs Program and the Lorne Trottier
Chair in Astrophysics and Cosmology.

%%% \bibliographystyle{apj}
%%% \bibliography{journals1,modrefs,psrrefs,crossrefs,extrarefs}
%%% % \begin{thebibliography}{62}
%%% % \expandafter\ifx\csname natexlab\endcsname\relax\def\natexlab#1{#1}\fi
%%% % \bibitem[{Alpar ... 
%%% % \end{thebibliography}

%% --------------------------------------------------------
%% FIGURES/PLOTS
%% --------------------------------------------------------
%% ** OBSERVATIONS
%% 1 plot-obslen.ps
%% 2 plot-obstime.ps
%% ** TIMING
%% 3 plot-timing.ps
%% 4 plot-flare1timing.ps
%% 5 plot-flare2timing.ps
%% 6 plot-glitch.ps
%% ** PULSE PROFILE
%% 7 plot-profiles.ps
%% 8 plot-chi.ps
%% 9 plot-twoprofiles.ps
%% ** PULSED FLUX
%% 10 plot-flux.ps
%% 11 plot-simulations.ps
%% ** BURSTS
%% 12 plot-burstlc.ps
%% 13 plot-spectrum.ps
%% 14 plot-bursts.ps
%% DISCUSSION
%% 15 plot-linearfit.ps
%% 16 plot-compare.ps
%% -------------------------------------------------------- 1
\clearpage
%% -------------------------------------------------------- 2
\clearpage
%% -------------------------------------------------------- 3 BOLD
\clearpage
%% -------------------------------------------------------- 4 BOLD
\clearpage
%% -------------------------------------------------------- 5 BOLD
\clearpage
%% -------------------------------------------------------- 6 BOLD
\clearpage
%% -------------------------------------------------------- 7
\clearpage
%% -------------------------------------------------------- 8
\clearpage
%% -------------------------------------------------------- 9
\clearpage
%% -------------------------------------------------------- 10 BOLD
\clearpage
%% -------------------------------------------------------- 11 BOLD
%%%% \clearpage
%%%% \begin{figure*}
%%%% \begin{center}
%%%% %\includegraphics[scale=0.85]{/homes/loiseau/xte/1048-collection-RESULTS/plot-simulations.ps}
%%%% \includegraphics[scale=0.85]{f11.eps}
%%%% \caption
%%%% {A comparison of {\emph{RXTE}} and {\emph{CXO}} spectral changes. 
%%%% {\emph{Top:}} Small crosses: {\emph{RXTE}} hardness ratio computed from the
%%%% pulsed flux in the energy range (4$-$10~keV)/(2$-$4~keV). Green circles:
%%%% hardness ratio computed from simulated {\emph{RXTE}} pulsed fluxes
%%%% at the epochs of five {\emph{CXO}} observations (see Section~\ref{sec:cxo}
%%%% for details). {\emph{Bottom:}} Pulsed fractions for the 5 {\emph{CXO}}
%%%% obsevations surrounding the onset of the third flare.
%%%% {\emph{Both Panels:}} the solid line indicates the onset of the flare.
%%%% \label{plot-simulations}}
%%%% \end{center}
%%%% \end{figure*}
%% -------------------------------------------------------- 12 now 11
\clearpage
%% -------------------------------------------------------- 13 BOLD now 12
\clearpage
%% -------------------------------------------------------- 14 BOLD now 13
\clearpage
%% -------------------------------------------------------- 15 BOLD now 14
%% -------------------------------------------------------- 16 BOLD now 15
\clearpage

%% --------------------------------------------------------
%% --------------------------------------------------------
%% --------------------------------------------------------
%% --------------------------------------------------------
%% --------------------------------------------------------
%% --------------------------------------------------------
%% -------------------------------------------------------- tableglitch
\clearpage
\begin{deluxetable}{lc}
\tablewidth{220pt}
\tablecaption
{
Local ephemeris of 1E~1048.1$-$5937 near the 2007 glitch\tablenotemark{a}
\label{tableglitch}
}
\tablehead
{
Parameter &
Value
}
\startdata
MJD range & 54164.545$-$54202.475 \\
TOAs & 21 \\
Epoch (MJD) & 54185.912956 \\ 
$\nu$ (s$^{-1}$) & 0.15484969(6) \\
$\dot{\nu}$ (s$^{-2}$) & $-$8.2(5)$\times$10$^{-13}$ \\
Glitch Epoch (MJD) & 54185.912956 \\
$\Delta\nu$ (s$^{-1}$) & 2.52(3)$\times$10$^{-6}$ \\
$\Delta\dot{\nu}$ (s$^{-2}$) & 6(4)$\times$10$^{-14}$ \\
RMS residual (phase) & 0.0098 \\
\enddata
\tablenotetext{a} {Numbers in parentheses are \texttt{TEMPO}-reported 1$\sigma$
uncertainties. }
\end{deluxetable}

%% --------------------------------------------------------
%% --------------------------------------------------------
%% --------------------------------------------------------
%% --------------------------------------------------------
%% --------------------------------------------------------
%% --------------------------------------------------------
%% -------------------------------------------------------- tableburst
\clearpage
\begin{deluxetable}{lc}
\tablewidth{300pt}
\tablecolumns{2}
%%% \tablewidth{\columnwidth}
\tablecaption{Burst Timing and Spectral Properties \label{table:burst}}
\tablehead{\multicolumn{2}{c}{Temporal Properties}  }
\startdata
Burst day (MJD) &  54218\\
Burst start time (fraction of day) & 0.578621(6)  \\
%%% Below is for an exponential model
%%% Burst rise time, $t_r$ (ms) & 509$^{+150}_{-114}$  \\
%%% Below is for a linear  model  HERE
Burst rise time, $t_r$ (ms) & 955$^{+80}_{-115}$ \\
Burst duration, $\tnin$ (s) & 111.2$^{+26}_{-19}$ \\
\cutinhead{Fluxes and Fluences} 
$\tnin$ fluence\tablenotemark{a}$\;$ (counts/PCU) & 445$\pm$15 \\
$\tnin$ fluence\tablenotemark{a}$\;$ ($\times 10^{-10}~\mathrm{erg~cm}^{-2}$) & 68.9$\pm$2.3 \\
Peak flux for 64~ms\tablenotemark{a}$\;$ ($\times 10^{-10}~\mathrm{erg~s^{-1}~cm}^{-2}$)    & 24.2$\pm$5.4   \\
%%% Peak flux for exponential rise time
%%% Peak flux for $t_r$~ms\tablenotemark{a} ($\times 10^{-10}~\mathrm{erg~s^{-1}~cm}^{-2}$) & 16.7$\pm$1.5  \\
%%% Peak flux foe linear rise time
Peak flux for $t_r$~ms\tablenotemark{a}$\;$ ($\times 10^{-10}~\mathrm{erg~s^{-1}~cm}^{-2}$) &  15.9$\pm$1.1 \\
\cutinhead{Spectral Properties} 
\underline{Power law:} & \\
Power law index & 0.37$^{+0.20}_{-0.19}$  \\
Unabsorbed power law flux ($\times 10^{-11}~\mathrm{erg~s^{-1}~cm^{-2}}$)  & 4.64$^{+0.44}_{-0.44}$  \\
Reduced $\chi^2$/degrees of freedom & 1.80/18  \\
 & \\
\underline{Blackbody:} &  \\
$kT$ (keV) & 4.9$^{+0.7}_{-0.6}$  \\
Unabsorbed Blackbody flux ($\times 10^{-11}~\mathrm{erg~s^{-1}~cm^{-2}}$) & 5.01$^{+1.32}_{-1.34}$ \\
Blackbody Radius\tablenotemark{b} (km) & 0.014$^{+0.006}_{-0.004}$  \\
Reduced $\chi^2$/degrees of freedom & 1.20/18 \\
\enddata
\tablenotetext{a}{Fluxes and fluences are calculated in the 2--20~keV band.}
\tablenotetext{b}{Assuming a distance of 9.0~kpc to the source \citep{dv06ax}.}
\end{deluxetable}

%% --------------------------------------------------------
%% --------------------------------------------------------
%% --------------------------------------------------------
%% --------------------------------------------------------
%% --------------------------------------------------------
%% --------------------------------------------------------
%% -------------------------------------------------------- tableline
\clearpage
\begin{deluxetable}{lccccccc}
\tablewidth{\textwidth}
\tablecolumns{8}
%%% \tablewidth{\columnwidth}
\tablecaption{Spectral Fit to the First 3 seconds of the New Burst \label{table:line}}
\tablehead{\colhead{Parameter} & \multicolumn{3}{r}{Value}}
\startdata 
$$ & Power-Law && Power-Law && Blackbody && Blackbody  \\
$$ & $$        && + Gaussian && $$       && +  Gaussian \\
\cline{2-2} \cline{4-4} \cline{6-6} \cline{8-8}\\
Index/Temperature (keV)& $\sim$0.44 &&  0.71$^{+0.28}_{-0.23}$  && 6.12$^{+0.80}_{-0.66}$   &&  5.26$^{+0.99}_{-0.99}$       \\ 
Continuum Flux\tablenotemark{a}$\;$ & 5.63$^{+0.46}_{-2.65}$  && 4.49$^{+0.60}_{-0.68}$ &&  6.55$^{+0.34}_{-0.77}$  && 5.44$^{+1.40}_{-1.24}$\\
Line Energy (keV) &  \nodata && 14.61$^{+0.50}_{-0.52}$   && \nodata && 14.88$^{+0.60}_{-0.57}$    \\
Line Width (keV) & \nodata &&  1.81$^{+0.73}_{-0.60}$ && \nodata && 1.35$^{+0.68}_{-0.83}$     \\
Line Flux\tablenotemark{a}$\;$ ($\times 10^{-10}~\mathrm{erg~s^{-1}~cm^{-2}}$) & \nodata && 2.70$^{+0.74}_{-0.79}$  &&  \nodata  &&  1.91$^{+1.00}_{-0.75}$ \\
Reduced $\chi^2$/degrees of freedom & 2.49/11 && 1.08/8 && 1.71/11  && 1.04/8   \\ 
\enddata
\tablecaption{\label{table:line}}
\tablenotetext{a}{Fluxes are unabsorbed and calculated in the
2--20~keV band, in units of $\times 10^{-10}~\mathrm{erg~s^{-1}~cm^{-2}}$.}
\end{deluxetable}
%%%
\clearpage
%% --------------------------------------------------------
%%% \end{document}
%% --------------------------------------------------------

%%% \bibliographystyle{apj}
%%% \bibliography{journals1,modrefs,psrrefs,crossrefs,extrarefs}
%%% % \begin{thebibliography}{62}
%%% % \expandafter\ifx\csname natexlab\endcsname\relax\def\natexlab#1{#1}\fi
%%% % \bibitem[{Alpar {\ldots}  
%%% % \end{thebibliography}

\bibliographystyle{apj}

%% --------------------------------------------------------
\end{document}